\documentclass[reprint,aps,prl,longbibliography]{revtex4-2}

\usepackage{epsfig,amssymb,amsmath,txfonts,hyperref,float} 
\usepackage{xcolor}
\usepackage{graphicx}
\usepackage[export]{adjustbox}
\usepackage{multirow}
\usepackage{pdfpages} 

\makeatletter 
\AtBeginDocument{\let\LS@rot\@undefined}
\makeatother

\begin{document}

\title{External magnetic field suppression of carbon diffusion in iron}

\author{Luke J. Wirth}
\affiliation{Department of Materials Science and Engineering, University of Illinois, Urbana-Champaign, Illinois 61801, USA}

\author{Dallas R. Trinkle}
\email{dtrinkle@illinois.edu}
\affiliation{Department of Materials Science and Engineering, University of Illinois, Urbana-Champaign, Illinois 61801, USA}

\date{\today}

\begin{abstract}

External magnetic fields reduce diffusion of carbon in BCC iron, but the physical mechanism is not understood. Using DFT calculations with magnetic moments sampled from a Heisenberg model, we calculate diffusivities of carbon in iron at high temperatures and with field. Our model reproduces the measured suppression of diffusivity from field. We find that increasing magnetic disorder flattens the electron density of states compared with the ferromagnetic case, which distorts the octahedral cages around carbon, lowering the activation barrier to diffusion; an applied field reverses these trends.

\end{abstract}

\maketitle

Controlling carbon diffusion in BCC iron is fundamental to processing and designing steel alloys. 
Experiments have found that different activation barriers and prefactors are necessary to describe zero-field carbon diffusion in BCC iron at low \cite{daSilva1976} compared to high temperatures \cite{Stanley1949}, with the prefactor increasing and the activation energy barrier decreasing approaching the Curie temperature.
In addition, external magnetic fields affect carbon diffusion in iron, despite carbon being a nonmagnetic element: Fujii and Tsurekawa observed suppression of carbon diffusion through BCC Fe parallel to externally applied magnetic fields \cite{Fujii2011}. Similarly, field effects on diffusion-controlled processes have been observed in nonmagnetic metal systems like aluminum alloys \cite{Kesler2025}. In BCC Fe, it has been hypothesized that the partial magnetic order induced by application of the field raises the activation energy barrier to carbon transitions \cite{Ruch1976}. 
However, this hypothesis lacks a physical connection between the magnetic ordering of Fe atoms and changes in carbon diffusion.

Various phenomenological models have been proposed to explain the experimental observations, but none provide a comprehensive physical explanation that applies to all high-temperature and high-field cases. Fujii and Tsurekawa \cite{Fujii2011} referred to McLellan's dual-occupancy model (DOM) for interstitial diffusion \cite{McLellan1965} and suggested that occupancy of tetrahedral sites decreases under field, reducing the ability of carbon to hop along tetrahedral-to-tetrahedral diffusion pathways. However, DFT studies showed that the tetrahedral C site in ferromagnetic BCC Fe is a saddle point and serves only as a transition state between octahedral equilibrium sites \cite{Jiang2003, Domain2004}; this is also true in the disordered local moment (DLM) paramagnetic case \footnote{See Supplemental Material at [URL will be inserted by publisher] for details on saddle point verification, how Zeeman fields affect the 0 K activation barrier, magnetic model and DFT calculation parameters, elastic dipole details, and activation energy statistics.}. Ruch \textit{et al.} \cite{Ruch1976}, following earlier work by Girifalco \cite{Girifalco1962} and Wuttig \cite{Wuttig1971} on non-Arrhenius diffusion generally seen in ferromagnets, offered a phenomenological explanation that the activation energy barrier $Q$ decreases with temperature-induced magnetic disorder according to $Q(T) = Q_\text{PM} + (Q_\text{FM}-Q_\text{PM}) S^2(T)$, where $Q_\text{PM}$ is the paramagnetic barrier in the infinite temperature limit, $Q_\text{FM}$ is the ferromagnetic barrier, and $S(T)$ is the magnetic saturation. Farraro and McLellan later concluded that the magnetic argument of Wuttig offers a more valid description of carbon in BCC Fe than the DOM, which overestimates diffusivity in the high-temperature BCC $\delta$-Fe phase \cite{Farraro1979}. Extension of the magnetic model to also account for field-induced magnetic order would allow for a consistent explanation of both temperature and field.

Previous computational efforts have failed to accurately predict high-temperature diffusion of carbon in BCC Fe and the suppression of diffusion under field. DFT studies of carbon diffusion in BCC Fe report activation energy barriers of 0.87 \cite{Jiang2003} and 0.92 eV \cite{Domain2004} at 0 K. These values are compatible with Arrhenius descriptions near the low-temperature limit \cite{Wert1950, daSilva1976} but lead to underestimates of high-temperature diffusivity \cite{Stanley1949}. Field effects at 0 K can be studied by applying a Zeeman splitting energy within DFT calculations \cite{Bousquet2011}, but the response of BCC Fe to Zeeman fields has not been published. We performed this calculation and found that a field on the order of 1000 T would be needed to increase the activation energy barrier by about 0.1 eV.
At higher temperatures, several general approaches exist for modeling spin fluctuations: DLM models can describe the fully paramagnetic case \cite{Neugebauer2014PRL}, partial DLM models can describe finite-temperature magnetic systems with unequal amounts of spin-up and spin-down moments \cite{Ruban2012}, or surrogate models can prepare noncollinear spin configurations expected at intermediate temperatures where short-range magnetic ordering persists even as long-range order is lost \cite{Neugebauer2014PRL,Kormann2011}. Noncollinear effects can be particularly important at or near the Curie temperature, where their contributions to magnetic disorder can dominate collinear spin-flip contributions \cite{Abrikosov2016}.
The time-scale separation between atomic motion and rapid spin fluctuations means that atoms experience average spin environments \cite{Kormann2012}. Spin-space averaging (SSA) computes the forces and energies as an average over multiple magnetic samples, allowing relaxed geometries to be found in the presence of magnetic disorder. Previous studies used SSA with DLM environments to relax vacancies and obtain vacancy migration barriers in BCC Fe \cite{Hegde2020} and to relax octahedral carbon in BCC Fe \cite{Alling2018}. Hegde and coworkers \cite{Hegde2021} modeled Mn diffusion in BCC Fe at finite temperatures by interpolating between FM and DLM-SSA limits with methods including the Ruch model, and by using a parameterized effective interaction model that was previously developed to study high-temperature self-diffusion and Cu diffusion in BCC Fe \cite{Schneider2020}. However, DLM configurations inherently cannot account for the presence of an external field.

We use Monte Carlo sampling of a Heisenberg model with and without external field for SSA with DFT at finite temperature to compute temperature- and field-dependent diffusivity of carbon in BCC iron. Sampled moments from Monte Carlo simulations serve as constraint directions for DFT calculations, which yield SSA forces that account for longitudinal moment fluctuations and geometric distortion around carbon. We relax octahedral and tetrahedral carbon with and without field at finite temperature to compute the average energy of each site. As the local alignment of iron spins increases, the barrier for carbon diffusion also increases with distortion of the neighboring environment. This first-principles prediction of carbon diffusion under field matches experiments, and elucidates the physical mechanism behind the suppression of carbon diffusion under magnetic field.

We compute the diffusivity $D$ at magnetic field magnitude $B_\text{ext}$ and temperature $T$ using SSA activation barriers $Q(B_\text{ext},T)$,
\begin{equation}
\label{eq:D}
    D(B_\text{ext}, T) = \frac{1}{6}a_0^2(T)\nu^*(T)f_\text{C}(T)\exp\bigg(-\frac{Q(B_\text{ext},T)}{k_\text{B}T}\bigg),
\end{equation}
where $a_0$ is the lattice constant, $\nu^*$ is the attempt frequency, $f_\text{C}$ is a correlation factor, and $k_\text{B}$ is the Boltzmann constant. Most of the computational effort lies in calculating $Q(B_\text{ext},T)$, which is the difference between the average energies of the tetrahedral transition state and octahedral equilibrium state over all samples from a set of conditions. Our calculations use the 0 K equilibrium volume, so we correct the energies with $\frac{1}{3}\text{Tr}(\Delta\overline{P}_{ij})\frac{V(T)}{V_0}$, where $\Delta\overline{P}_{ij}$ is the tetrahedral–octahedral difference in the DFT-computed derivative of the elastic energy with respect to volumetric strain, ${V(T)}$ is the empirical thermal volumetric expected for BCC Fe at $T$, and $V_0$ is the volume at 0 K \cite{Acet1994}. Thermal expansion also modifies $a_0^2(T)$ in Eq. \ref{eq:D}. We use Vineyard's model \cite{Vineyard1957} with the hopping atom approximation \cite{Garnier2013}, which is appropriate given the difference in mass between carbon and iron by nearly a factor of five, and find $\nu^*(0\;\text{K})$ to be 10.6 THz. To account for the temperature dependence of $\nu^*$, we introduce an empirical softening factor based on K\"ormann \textit{et al.}'s finding that magnetic disorder softens phonons in BCC Fe, with maximal softening appearing at the N-point phonon \cite{Neugebauer2014PRL}. This is relevant to carbon diffusion that takes place along $\langle$100$\rangle$ pathways, and based on the magnitude of observed softening we expect temperature-induced magnetic disorder to suppress vibrational contributions to the diffusion prefactor approximately by a factor of two; therefore, $\nu^*(T \gtrsim T_\text{C}) =$ 5.3 THz, where $T_\text{C}$ is the Curie temperature. 
Finally, following the molecular dynamics study of Tapasa \textit{et al.}, we assume that $f_\text{C} = 0.66$ at temperatures below 1200 K, accounting for the tendency of carbon hops that cross the activation barrier to recross it before equilibrating \cite{Osetsky2006}.

Monte Carlo simulations in large BCC Fe supercells use the Metropolis algorithm and a Heisenberg Hamiltonian with an exchange parameter tuned to yield the zero-field empirical Curie temperature of 1043 K. Simulations take place at four sets of conditions: at 1043 K with and without a 6 T field, to compare with the diffusion experiment conducted by Fujii and Tsurekawa \cite{Fujii2011}; at 986 K without field where the same net magnetization is observed as in the previous 6 T case, to assess the importance of the field independent of the net magnetization; and in the totally random disordered local moment (DLM) case, which provides information on the high-temperature limit to complement the low-temperature ferromagnetic case. Spin-sampling involves taking 25 sets of magnetic moments from a 54-atom cubic region positioned within the larger supercell, with each set of samples separated from one another by the autocorrelation time to ensure sample independence. The directions of these moments then serve as constraints on Fe atoms in Fe$_{54}$C supercells, which we impose using the algorithm of Ma and Dudarev \cite{Ma2016}, within DFT calculations using \textsc{vasp} \cite{Kresse1993, Kresse1994, Kresse1996, Kresse1996B, Kresse1999}. We relax each geometry by displacing atoms along the symmetrized, averaged DFT forces calculated in a corresponding set of Fe$_{54}$C supercells. The carbon atom sits in the center of each sampled set of spins, and does not move during relaxation. Force symmetrization effectively increases our number of 25 samples by factors of eight and four to 200 and 100 for the octahedral and tetrahedral configurations respectively. To facilitate relaxation, we generate force constant matrices for the octahedral and tetrahedral configurations by displacing the carbon atom by 0.01 $\AA$ in separate ferromagnetic calculations. We compute lattice Green's functions to efficiently relax iron atoms around carbon from the symmetrized SSA forces. This requires less than 10 ionic updates to converge symmetrized forces to below 20 meV/$\AA$. The relaxation can be sped up by taking fractional steps due to the ferromagnetic 0 K case being stiffer than those with magnetic disorder present. Previous SSA works from the literature discuss different ways of handling symmetrized forces \cite{Hegde2020,Alling2018}. 

\begin{figure}[htbp]
\includegraphics[width=3.375in]{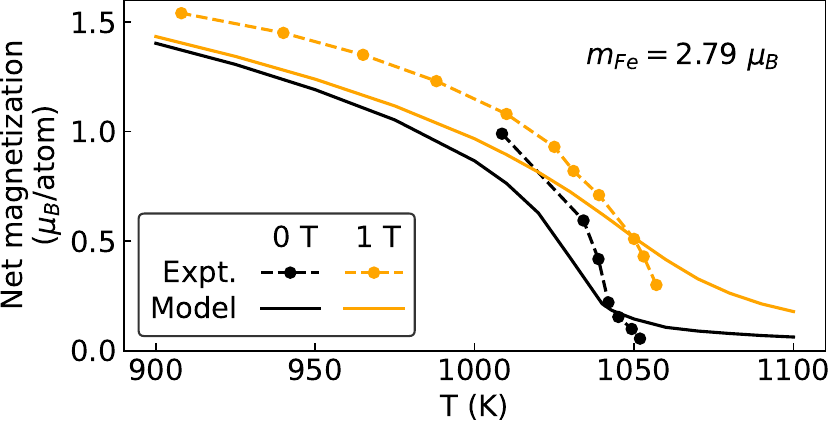}
\caption{Net magnetization in BCC Fe with temperature and field in experiment and with a parameterized Heisenberg model. Dashed lines are experimentally observed values in the zero-field \cite{Potter1934} and externally applied field \cite{Crangle1971}, with solid lines for the model.}
\label{fig:M}
\end{figure}

Fig.\ \ref{fig:M} shows how our Heisenberg model reproduces the experimentally observed temperature-dependence of the net magnetization of iron with \cite{Crangle1971} and without \cite{Potter1934} the presence of an external magnetic field near the Curie temperature. We use a periodic cell with length $L = 32$ containing $2^{16}=65,536$ Fe atoms. An exchange interaction parameter $J=43.2\text{ meV}$ reproduces the experimental Curie temperature of 1043 K; this parameter would vary slightly for larger cells\cite{Lundow2009}. We tune the magnetic moment magnitude of each Fe atom in the Monte Carlo simulations to $2.79 \mu_\text{B}$ to reproduce the response to fields of 0.5, 1.0, and 1.5 T. This magnitude is significantly greater than below room temperature value of 2.20 $\mu_\text{B}$ \cite{Crangle1971}. However, measurements of the paramagnetic susceptibility by Arajs and Miller found that from 1100--1180 K, BCC Fe responds to external fields with local moments of magnitude 3.13 $\mu_\text{B}$ \cite{Arajs1960}, suggesting that 2.79 is reasonable. The good agreement between model and experiment shows that the Heisenberg model accurately describes both field- and temperature-dependent long-range order in BCC Fe.

\begin{figure}[htbp]
\includegraphics[width=3.375in]{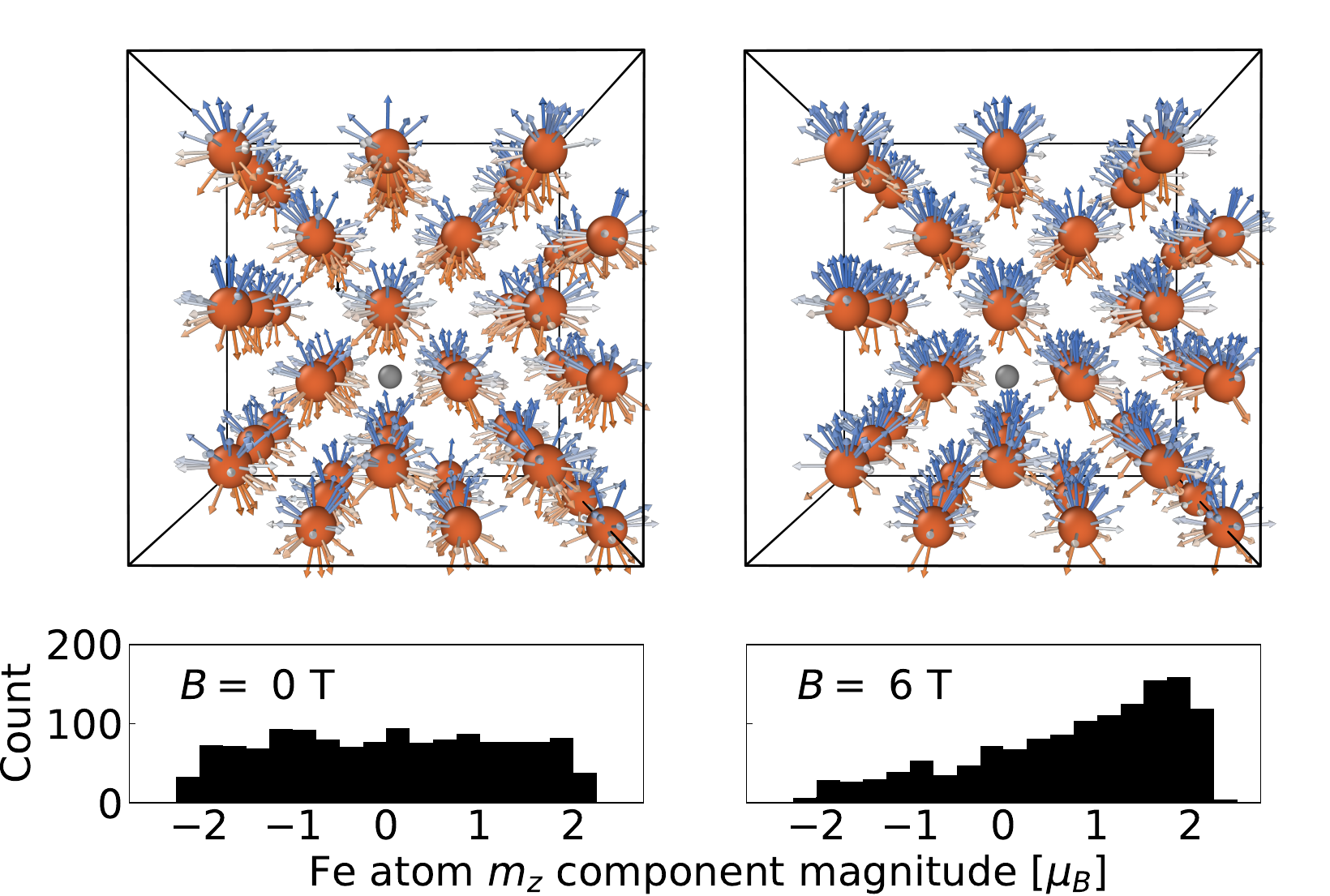}
\caption{Magnetic environments from Heisenberg model sampling in a BCC Fe$_{54}$C supercell at the Curie temperature of 1043 K in the zero-field case (left) and with an externally applied 6 T magnetic field (right). All 25 sets of moments sampled for DFT calculations are visualized simultaneously, with moments color-coded according to their component parallel to the field direction. Histograms below each plot show the distributions of component magnitudes after DFT relaxation.}
\label{fig:moments}
\end{figure}
    
Fig.\ \ref{fig:moments} illustrates the Fe$_{54}$C supercell used for noncollinear DFT calculations, and shows how the magnetic moment distributions generated by our models vary with and without an external magnetic field. Visualizations depict all 25 sets of sampled spin-spaces simultaneously; the tendency of moments to align with the 6 T field, but not to the point of saturation, is visually apparent. Histograms in the bottom panels depict distributions of the components of each moment parallel to the applied field. Fluctuations of moment magnitude enable corrections to be made to the configurations generated by the surrogate model, particularly by allowing for relaxation of moments in the strained geometry around carbon, where it is known that suppressed moments exist at 0 K \cite{Domain2004}. A tradeoff of this approach is that since the DFT calculations take place at 0 K and we don't constrain moment magnitudes, the moments here don't approach the high magnitudes expected near $T_\text{C}$. But, because the increased net ordering that those moments imposed within the surrogate model is present in the set of constraint directions, we expect that the energy differences between tetrahedral and octahedral configurations within these sets of $\sim$2.20 $\mu_\text{B}$ moments are similar to those we would see with $\sim$2.79 $\mu_\text{B}$ moments. The significantly different average environments seen with and without field lead to differences in the activation energies that help to describe why field-induced order suppresses diffusion of carbon. 

\begin{figure}[htbp]
\includegraphics[width=3.375in]{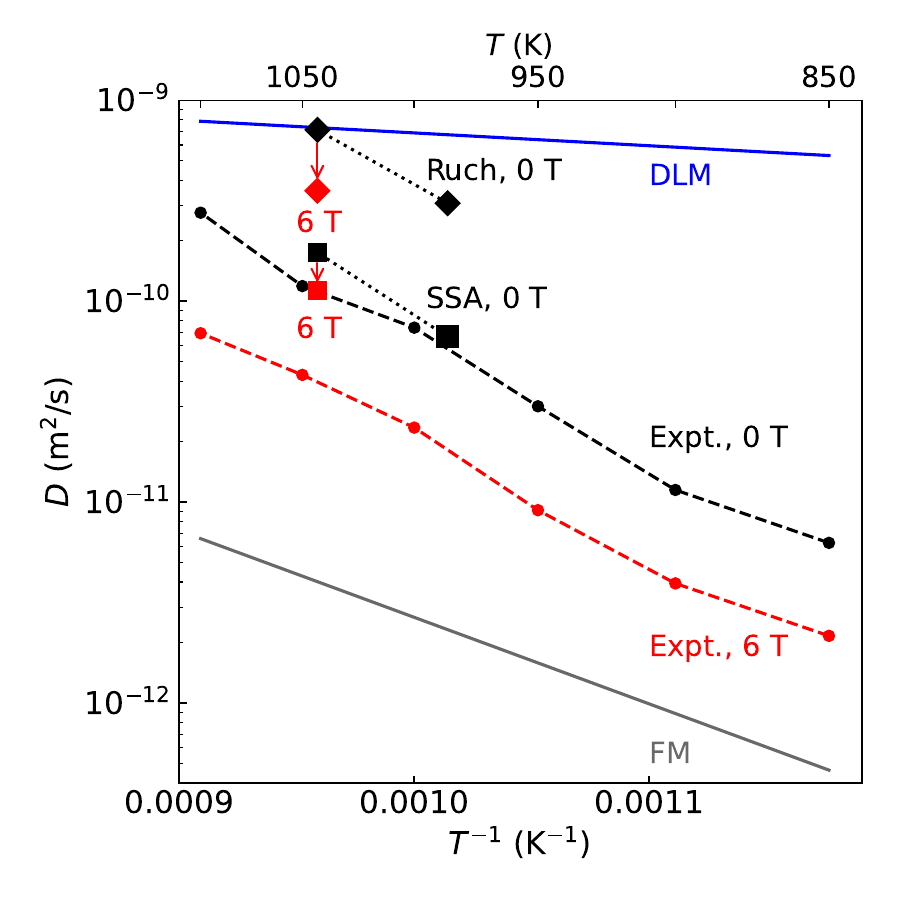}
\caption{Diffusivity of carbon in BCC Fe at conditions of interest as modeled by spin-space averaged (SSA) calculations. The SSA models closely agree with experimental observations made by Fujii and Tsurekawa \cite{Fujii2011}, and are notably more accurate than predictions that can be obtained by simple ferromagnetic (FM) or disordered local moment (DLM) models, or Ruch model interpolations between the two.}
\label{fig:D}
\end{figure}
    
Fig.\ \ref{fig:D} shows how spin-space averaged (SSA) models of diffusivity of carbon in BCC Fe at temperatures and fields of interest agree well with experimental measurements \cite{Fujii2011}, while the paramagnetic (DLM) or ferromagnetic (FM) models disagree. The DLM and FM models use activation barriers calculated in the randomly disordered and completely aligned cases while also taking into account the finite-temperature effects included in Eq. \ref{eq:D}. These endpoints provide diffusivities that span several orders of magnitude, but with neither matching the experiment. At $T_\text{C}$, net magnetic order is near zero, so the Ruch model predicts the DLM barrier in the zero-field case. Ruch model predictions using saturations from Monte Carlo simulations at lower finite temperatures also overestimate diffusivity; for example, at 986 K, $S^2$ is just $\sim$12\% of the FM value and the estimated $Q$ remains close to the DLM value. In contrast, zero-field SSA diffusion coefficients calculated at 986 and 1043 K do agree well with the trend experimentally observed by Fujii and Tsurekawa \cite{Fujii2011}, where the SSA model successfully captures the effects of magnetic short range order on $Q$. The SSA methodology also reproduces the observed suppression of diffusivity by the 6 T field, where the expected net magnetization is $\sim$35\% of the 0 K case. It is possible that an even better match with the experimental field response could be attained by including iron-carbon interactions and longitudinal moment fluctuations during moment direction sampling, though the agreement that we already see suggests that these terms would yield much smaller effects.

\begin{figure}[htbp]
\includegraphics[width=3.375in]{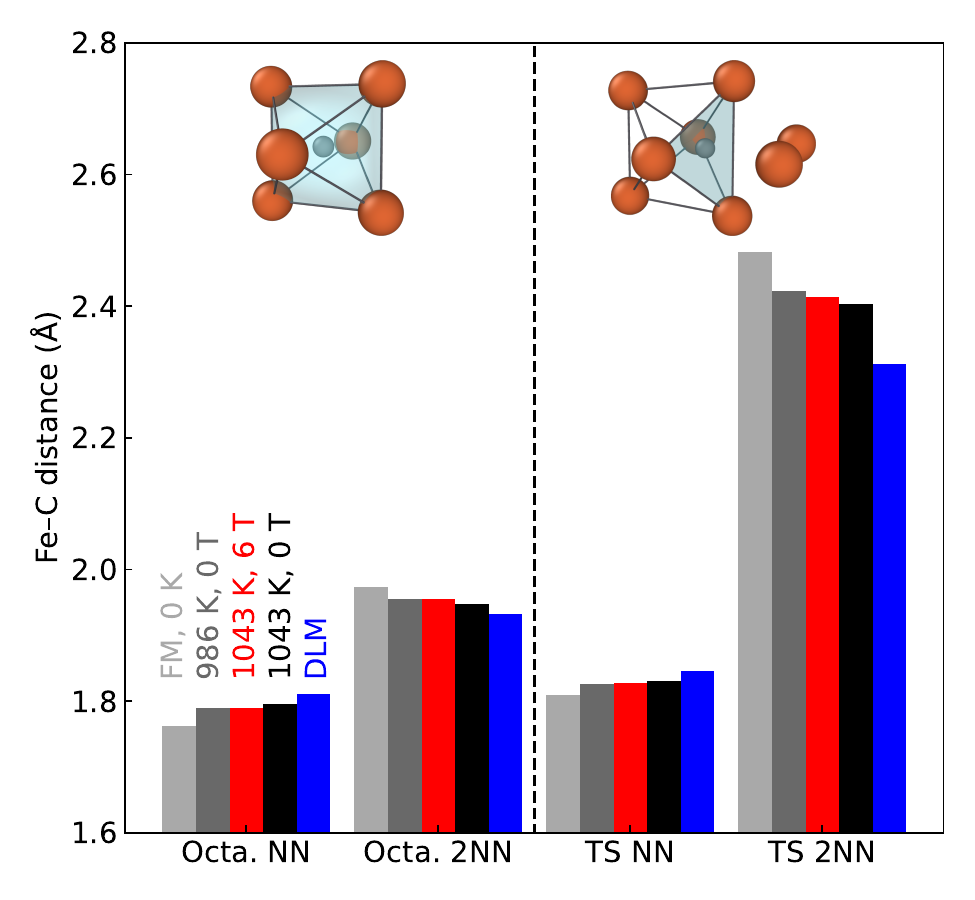}
\caption{Effect of magnetic ordering on Fe–C nearest-neighbor (NN) and second-nearest-neighbor (2NN) interatomic distances when carbon sits at $\langle$100$\rangle$ octahedral sites and tetrahedral transition states (TS). Geometric insets illustrate carbon and its NNs and 2NNs for each configuration. The shaded region in each panel indicates the octahedral or tetrahedral cage, while the black lines outline the octahedral cage in both panels.}
\label{fig:FeC}
\end{figure}

The SSA calculations show changes in the octahedral cage driven by field- or temperature-induced order, which affects the activation barrier for diffusion. The octahedral cage in BCC Fe has tetragonal symmetry, with two close neighbors and four next neighbors. The Fe–C interatomic distances in Fig.\ \ref{fig:FeC} show the octahedral cage surrounding carbon becoming more isotropic with magnetic disorder compared to the 0 K case which lowers the barrier to diffusion. The nearest neighbors expand while the second neighbors contract; the FM and DLM octahedral configurations agreeing with calculations of Gambino and Alling \cite{Alling2018}. The (986 K, 0 T) and (1043 K, 6 T) cases with similar net magnetizations have distances that are closer to one another than the (1043 K, 0 T) values, indicating that the geometric response to magnetization behaves the same way from decreased temperature or increased external field. Similarly, the elastic corrections due to expected thermal expansion that we apply to Eq. \ref{eq:D} vary with more compressive dipoles observed in the higher-magnetization conditions. The DLM distances demonstrate more of a deviation from the 0 K values than the finite-T ones, indicating that the short range order present in those cases has an effect on the geometries. This provides a qualitative explanation for the observed effect of the field reducing diffusivity as seen in Fig.\ \ref{fig:D} as well as the non-Arrhenius change in the zero-field activation barrier with temperature.

\begin{figure}[htbp]
\centering
\includegraphics[width=3.375in]{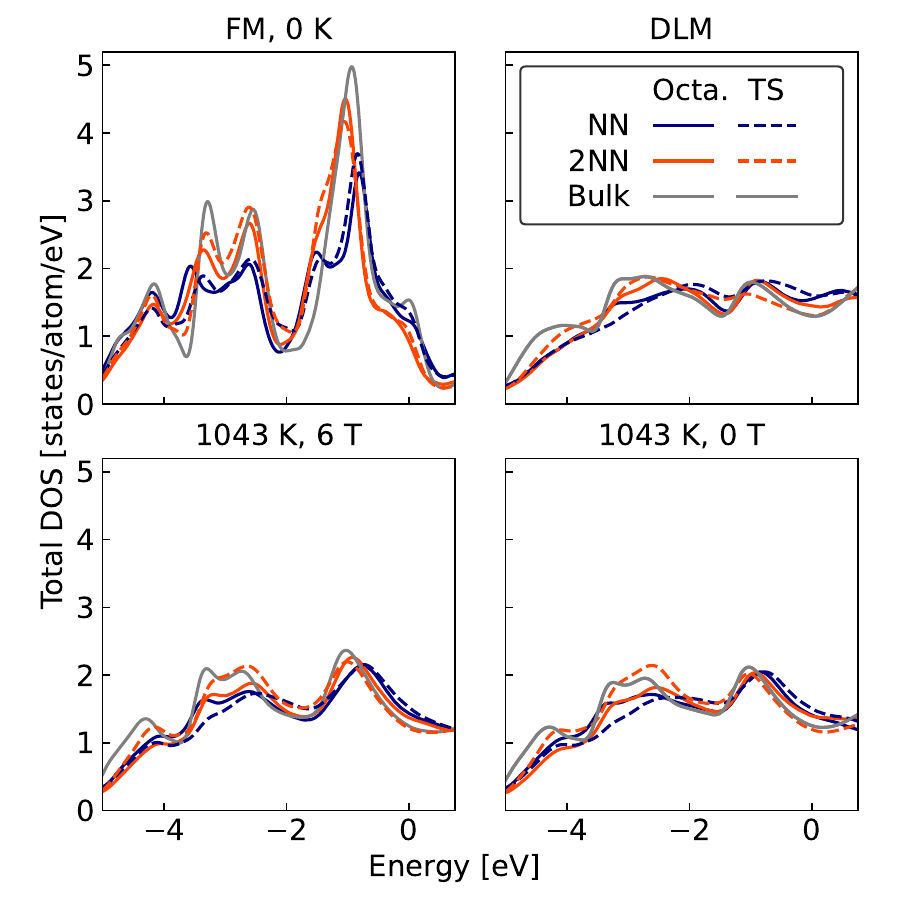}
\caption{Density of states plots for the nearest (NN) and second nearest (2NN) iron neighbors of carbon in the octahedral and tetrahedral configurations, with comparison made to bulk iron. Magnetic disorder increases counterclockwise from the upper-left ferromagnetic (FM) panel, with the disordered local moment (DLM) configuration having randomly oriented moments.}
\label{fig:DOS}
\end{figure}

Fig.\ \ref{fig:DOS} further helps to explain the quantitative findings by showing how magnetic disorder affects local density-of-states (LDOS) distributions for octahedral and tetrahedral Fe–C configurations, filling in the pseudogap seen at 0 K with higher DOS values just above the Fermi level. Pronounced peaks and valleys of the 0 K FM LDOS exist for bulk BCC Fe as well as neighbors of carbon, with an abundance of states existing just below the Fermi level before the pseudogap appears. Beginning with the FM case and moving counterclockwise through the panels of the figure, magnetic disorder fills in the pseudogap seen just above the Fermi level, and the total states become more evenly distributed across energies. This leads to more free-electron-like behavior in iron compared to the ordered case, which is likely why the tetragonal geometries around carbon become more isotropic at high temperatures or in the absence of an applied field.

First-principles calculations use spin-space averaging to quantify the effects of temperature and magnetic field on carbon diffusion in BCC Fe. The activation barrier increases with magnetic ordering imposed by either temperature or field. The simple Heisenberg model reproduces $T_\text{C}$ and the response to field, which allows sampling of the local spin environments that the Fe neighbors of C will see at higher temperatures and with field. Effects of short-range magnetic ordering at the Curie temperature cannot be described by referring to the ferromagnetic barrier calculated at 0 K, the fully random disordered local moment barrier, or an interpolation between the two. Densities of states indicate that electrons in BCC Fe behave more like free electrons as magnetic disorder increases. This causes the tetragonal geometries around carbon to become more isotropic, effectively opening the cages that contain carbon at equilibrium sites, lowering the barrier to diffusion. A mechanistic understanding of the influence of magnetic field on diffusion opens the possibility to design alloys that leverage this effect for improved properties or processing. Moreover, our quantitative approach can be applied to study of diffusion under magnetic fields in other materials like aluminum, and input for modeling diffusion controlled phase transitions under fields.

\section{ACKNOWLEDGMENTS}

This material is based upon work supported by the U.S. Department of Energy’s Office of Energy Efficiency and Renewable Energy (EERE) under the Advanced Manufacturing Office award number DE-EE0009131. The research was performed using computational resources sponsored by the EERE and located at the National Renewable Energy Laboratory. Atomic geometry visualizations used \textsc{ovito} \cite{Stukowski2010}. The data is available at the Materials Data Facility \cite{Blaszik2016, Blaszik2019}, doi:10.18126/fttq-w045 \cite{Wirth2025data}.


\begin{thebibliography}{46}%
\makeatletter
\providecommand \@ifxundefined [1]{%
 \@ifx{#1\undefined}
}%
\providecommand \@ifnum [1]{%
 \ifnum #1\expandafter \@firstoftwo
 \else \expandafter \@secondoftwo
 \fi
}%
\providecommand \@ifx [1]{%
 \ifx #1\expandafter \@firstoftwo
 \else \expandafter \@secondoftwo
 \fi
}%
\providecommand \natexlab [1]{#1}%
\providecommand \enquote  [1]{``#1''}%
\providecommand \bibnamefont  [1]{#1}%
\providecommand \bibfnamefont [1]{#1}%
\providecommand \citenamefont [1]{#1}%
\providecommand \href@noop [0]{\@secondoftwo}%
\providecommand \href [0]{\begingroup \@sanitize@url \@href}%
\providecommand \@href[1]{\@@startlink{#1}\@@href}%
\providecommand \@@href[1]{\endgroup#1\@@endlink}%
\providecommand \@sanitize@url [0]{\catcode `\\12\catcode `\$12\catcode
  `\&12\catcode `\#12\catcode `\^12\catcode `\_12\catcode `\%12\relax}%
\providecommand \@@startlink[1]{}%
\providecommand \@@endlink[0]{}%
\providecommand \url  [0]{\begingroup\@sanitize@url \@url }%
\providecommand \@url [1]{\endgroup\@href {#1}{\urlprefix }}%
\providecommand \urlprefix  [0]{URL }%
\providecommand \Eprint [0]{\href }%
\providecommand \doibase [0]{https://doi.org/}%
\providecommand \selectlanguage [0]{\@gobble}%
\providecommand \bibinfo  [0]{\@secondoftwo}%
\providecommand \bibfield  [0]{\@secondoftwo}%
\providecommand \translation [1]{[#1]}%
\providecommand \BibitemOpen [0]{}%
\providecommand \bibitemStop [0]{}%
\providecommand \bibitemNoStop [0]{.\EOS\space}%
\providecommand \EOS [0]{\spacefactor3000\relax}%
\providecommand \BibitemShut  [1]{\csname bibitem#1\endcsname}%
\let\auto@bib@innerbib\@empty
\bibitem [{\citenamefont {da~Silva}\ and\ \citenamefont
  {McLellan}(1976)}]{daSilva1976}%
  \BibitemOpen
  \bibfield  {author} {\bibinfo {author} {\bibfnamefont {J.~R.~G.}\
  \bibnamefont {da~Silva}}\ and\ \bibinfo {author} {\bibfnamefont {R.~B.}\
  \bibnamefont {McLellan}},\ }\bibfield  {title} {\bibinfo {title} {Diffusion
  of carbon and nitrogen in b.c.c. iron},\ }\href
  {https://doi.org/https://doi.org/10.1016/0025-5416(76)90229-9} {\bibfield
  {journal} {\bibinfo  {journal} {Mater. Sci. and Eng.}\ }\textbf {\bibinfo
  {volume} {26}},\ \bibinfo {pages} {83} (\bibinfo {year} {1976})}\BibitemShut
  {NoStop}%
\bibitem [{\citenamefont {Stanley}(1949)}]{Stanley1949}%
  \BibitemOpen
  \bibfield  {author} {\bibinfo {author} {\bibfnamefont {J.~K.}\ \bibnamefont
  {Stanley}},\ }\bibfield  {title} {\bibinfo {title} {The diffusion and
  solubility of carbon in alpha iron},\ }\href
  {https://doi.org/https://doi.org/10.1007/BF03398933} {\bibfield  {journal}
  {\bibinfo  {journal} {JOM}\ }\textbf {\bibinfo {volume} {1}},\ \bibinfo
  {pages} {752} (\bibinfo {year} {1949})}\BibitemShut {NoStop}%
\bibitem [{\citenamefont {Fujii}\ and\ \citenamefont
  {Tsurekawa}(2011)}]{Fujii2011}%
  \BibitemOpen
  \bibfield  {author} {\bibinfo {author} {\bibfnamefont {H.}~\bibnamefont
  {Fujii}}\ and\ \bibinfo {author} {\bibfnamefont {S.}~\bibnamefont
  {Tsurekawa}},\ }\bibfield  {title} {\bibinfo {title} {Diffusion of carbon in
  iron under magnetic fields},\ }\href
  {https://doi.org/http://dx.doi.org/10.1103/PhysRevB.83.054412} {\bibfield
  {journal} {\bibinfo  {journal} {Phys. Rev. B}\ }\textbf {\bibinfo {volume}
  {83}},\ \bibinfo {pages} {054412} (\bibinfo {year} {2011})}\BibitemShut
  {NoStop}%
\bibitem [{\citenamefont {Kesler}\ \emph {et~al.}(2025)\citenamefont {Kesler},
  \citenamefont {Thompson}, \citenamefont {Weiss},\ and\ \citenamefont
  {Manuel}}]{Kesler2025}%
  \BibitemOpen
  \bibfield  {author} {\bibinfo {author} {\bibfnamefont {M.~S.}\ \bibnamefont
  {Kesler}}, \bibinfo {author} {\bibfnamefont {M.~J.}\ \bibnamefont
  {Thompson}}, \bibinfo {author} {\bibfnamefont {D.}~\bibnamefont {Weiss}},\
  and\ \bibinfo {author} {\bibfnamefont {M.~V.}\ \bibnamefont {Manuel}},\
  }\bibfield  {title} {\bibinfo {title} {Simultaneously improving process
  efficiency and mechanical properties in aluminum alloys with applied magnetic
  fields},\ }in\ \href
  {https://doi.org/https://doi.org/10.1007/978-3-031-80676-6_32} {\emph
  {\bibinfo {booktitle} {Light Metals 2025}}},\ \bibinfo {editor} {edited by\
  \bibinfo {editor} {\bibfnamefont {L.}~\bibnamefont {Edwards}}}\ (\bibinfo
  {publisher} {Springer Nature Switzerland},\ \bibinfo {address} {Cham},\
  \bibinfo {year} {2025})\ p.\ \bibinfo {pages} {246–251}\BibitemShut
  {NoStop}%
\bibitem [{\citenamefont {Ruch}\ \emph {et~al.}(1976)\citenamefont {Ruch},
  \citenamefont {Sain}, \citenamefont {Yeh},\ and\ \citenamefont
  {Girifalco}}]{Ruch1976}%
  \BibitemOpen
  \bibfield  {author} {\bibinfo {author} {\bibfnamefont {L.}~\bibnamefont
  {Ruch}}, \bibinfo {author} {\bibfnamefont {D.~R.}\ \bibnamefont {Sain}},
  \bibinfo {author} {\bibfnamefont {H.~L.}\ \bibnamefont {Yeh}},\ and\ \bibinfo
  {author} {\bibfnamefont {L.}~\bibnamefont {Girifalco}},\ }\bibfield  {title}
  {\bibinfo {title} {Analysis of diffusion in ferromagnets},\ }\href
  {https://doi.org/https://doi.org/10.1016/0022-3697(76)90001-9} {\bibfield
  {journal} {\bibinfo  {journal} {J. Phys. Chem. Solids}\ }\textbf {\bibinfo
  {volume} {37}},\ \bibinfo {pages} {649} (\bibinfo {year} {1976})}\BibitemShut
  {NoStop}%
\bibitem [{\citenamefont {McLellan}\ \emph {et~al.}(1965)\citenamefont
  {McLellan}, \citenamefont {Rudee},\ and\ \citenamefont
  {Ishibachi}}]{McLellan1965}%
  \BibitemOpen
  \bibfield  {author} {\bibinfo {author} {\bibfnamefont {R.~B.}\ \bibnamefont
  {McLellan}}, \bibinfo {author} {\bibfnamefont {M.~L.}\ \bibnamefont
  {Rudee}},\ and\ \bibinfo {author} {\bibfnamefont {T.}~\bibnamefont
  {Ishibachi}},\ }\bibfield  {title} {\bibinfo {title} {The thermodynamics of
  dilute interstitial solid solutions with dual-site occupancy and its
  application to the diffusion of carbon in alpha iron},\ }\href@noop {}
  {\bibfield  {journal} {\bibinfo  {journal} {AIME Met. Soc. Trans.}\ }\textbf
  {\bibinfo {volume} {233}},\ \bibinfo {pages} {1938} (\bibinfo {year}
  {1965})}\BibitemShut {NoStop}%
\bibitem [{\citenamefont {Jiang}\ and\ \citenamefont
  {Carter}(2003)}]{Jiang2003}%
  \BibitemOpen
  \bibfield  {author} {\bibinfo {author} {\bibfnamefont {D.~E.}\ \bibnamefont
  {Jiang}}\ and\ \bibinfo {author} {\bibfnamefont {E.~A.}\ \bibnamefont
  {Carter}},\ }\bibfield  {title} {\bibinfo {title} {Carbon dissolution and
  diffusion in ferrite and austenite from first principles},\ }\href
  {https://doi.org/https://doi.org/10.1103/PhysRevB.67.214103} {\bibfield
  {journal} {\bibinfo  {journal} {Phys. Rev. B}\ }\textbf {\bibinfo {volume}
  {67}},\ \bibinfo {pages} {214103} (\bibinfo {year} {2003})}\BibitemShut
  {NoStop}%
\bibitem [{\citenamefont {Domain}\ \emph {et~al.}(2004)\citenamefont {Domain},
  \citenamefont {Becquart},\ and\ \citenamefont {Foct}}]{Domain2004}%
  \BibitemOpen
  \bibfield  {author} {\bibinfo {author} {\bibfnamefont {C.}~\bibnamefont
  {Domain}}, \bibinfo {author} {\bibfnamefont {C.~S.}\ \bibnamefont
  {Becquart}},\ and\ \bibinfo {author} {\bibfnamefont {J.}~\bibnamefont
  {Foct}},\ }\bibfield  {title} {\bibinfo {title} {Ab initio study of foreign
  interstitial atom {(C,N)} interactions with intrinsic point defects in
  {$\alpha$–Fe}},\ }\href
  {https://doi.org/https://doi.org/10.1103/PhysRevB.69.144112} {\bibfield
  {journal} {\bibinfo  {journal} {Phys. Rev. B}\ }\textbf {\bibinfo {volume}
  {69}},\ \bibinfo {pages} {144112} (\bibinfo {year} {2004})}\BibitemShut
  {NoStop}%
\bibitem [{Note1()}]{Note1}%
  \BibitemOpen
  \bibinfo {note} {See Supplemental Material at [URL will be inserted by
  publisher] for details on saddle point verification, how Zeeman fields affect
  the 0 K activation barrier, magnetic model and DFT calculation parameters,
  elastic dipole details, and activation energy statistics.}\BibitemShut
  {Stop}%
\bibitem [{\citenamefont {Girifalco}(1962)}]{Girifalco1962}%
  \BibitemOpen
  \bibfield  {author} {\bibinfo {author} {\bibfnamefont {L.~A.}\ \bibnamefont
  {Girifalco}},\ }\bibfield  {title} {\bibinfo {title} {Activation energy for
  diffusion in ferromagnetics},\ }\href
  {https://doi.org/https://doi.org/10.1016/0022-3697(62)90136-1} {\bibfield
  {journal} {\bibinfo  {journal} {J. Phys. Chem. Solids}\ }\textbf {\bibinfo
  {volume} {23}},\ \bibinfo {pages} {1171} (\bibinfo {year}
  {1962})}\BibitemShut {NoStop}%
\bibitem [{\citenamefont {Wuttig}(1971)}]{Wuttig1971}%
  \BibitemOpen
  \bibfield  {author} {\bibinfo {author} {\bibfnamefont {M.}~\bibnamefont
  {Wuttig}},\ }\bibfield  {title} {\bibinfo {title} {On the activation entropy
  of interstitial diffusion in b.c.c. iron},\ }\href
  {https://doi.org/https://doi.org/10.1016/0036-9748(71)90222-5} {\bibfield
  {journal} {\bibinfo  {journal} {Scripta Metall.}\ }\textbf {\bibinfo {volume}
  {5}},\ \bibinfo {pages} {33} (\bibinfo {year} {1971})}\BibitemShut {NoStop}%
\bibitem [{\citenamefont {Farraro}\ and\ \citenamefont
  {McLellan}(1978)}]{Farraro1979}%
  \BibitemOpen
  \bibfield  {author} {\bibinfo {author} {\bibfnamefont {R.}~\bibnamefont
  {Farraro}}\ and\ \bibinfo {author} {\bibfnamefont {R.~B.}\ \bibnamefont
  {McLellan}},\ }\bibfield  {title} {\bibinfo {title} {The diffusion of heavy
  interstitial solute atoms in body-centered cubic metals},\ }\href
  {https://doi.org/https://doi.org/10.1016/0025-5416(79)90169-1} {\bibfield
  {journal} {\bibinfo  {journal} {Mater. Sci. and Eng.}\ }\textbf {\bibinfo
  {volume} {39}},\ \bibinfo {pages} {47–56} (\bibinfo {year}
  {1978})}\BibitemShut {NoStop}%
\bibitem [{\citenamefont {Wert}(1950)}]{Wert1950}%
  \BibitemOpen
  \bibfield  {author} {\bibinfo {author} {\bibfnamefont {C.~A.}\ \bibnamefont
  {Wert}},\ }\bibfield  {title} {\bibinfo {title} {Diffusion coefficient of {C}
  in $\alpha$-iron},\ }\href
  {https://doi.org/https://doi.org/10.1103/PhysRev.79.601} {\bibfield
  {journal} {\bibinfo  {journal} {Phys. Rev.}\ }\textbf {\bibinfo {volume}
  {79}},\ \bibinfo {pages} {601} (\bibinfo {year} {1950})}\BibitemShut
  {NoStop}%
\bibitem [{\citenamefont {Bousquet}\ \emph {et~al.}(2011)\citenamefont
  {Bousquet}, \citenamefont {Spaldin},\ and\ \citenamefont
  {Delaney}}]{Bousquet2011}%
  \BibitemOpen
  \bibfield  {author} {\bibinfo {author} {\bibfnamefont {E.}~\bibnamefont
  {Bousquet}}, \bibinfo {author} {\bibfnamefont {N.~A.}\ \bibnamefont
  {Spaldin}},\ and\ \bibinfo {author} {\bibfnamefont {K.~T.}\ \bibnamefont
  {Delaney}},\ }\bibfield  {title} {\bibinfo {title} {Unexpectedly large
  electronic contribution to linear magnetoelectricity},\ }\href
  {https://doi.org/https://doi.org/10.1103/PhysRevLett.106.107202} {\bibfield
  {journal} {\bibinfo  {journal} {Phys. Rev. Lett.}\ }\textbf {\bibinfo
  {volume} {106}},\ \bibinfo {pages} {107202} (\bibinfo {year}
  {2011})}\BibitemShut {NoStop}%
\bibitem [{\citenamefont {K{\"o}rmann}\ \emph {et~al.}(2014)\citenamefont
  {K{\"o}rmann}, \citenamefont {Grabowski}, \citenamefont {Dutta},
  \citenamefont {Hickel}, \citenamefont {Fultz},\ and\ \citenamefont
  {Neugebauer}}]{Neugebauer2014PRL}%
  \BibitemOpen
  \bibfield  {author} {\bibinfo {author} {\bibfnamefont {F.}~\bibnamefont
  {K{\"o}rmann}}, \bibinfo {author} {\bibfnamefont {B.}~\bibnamefont
  {Grabowski}}, \bibinfo {author} {\bibfnamefont {B.}~\bibnamefont {Dutta}},
  \bibinfo {author} {\bibfnamefont {T.}~\bibnamefont {Hickel}}, \bibinfo
  {author} {\bibfnamefont {B.}~\bibnamefont {Fultz}},\ and\ \bibinfo {author}
  {\bibfnamefont {J.}~\bibnamefont {Neugebauer}},\ }\bibfield  {title}
  {\bibinfo {title} {Temperature dependent magnon-phonon coupling in bcc {Fe}
  from theory and experiment},\ }\href
  {https://doi.org/https://doi.org/10.1103/PhysRevLett.113.165503} {\bibfield
  {journal} {\bibinfo  {journal} {Phys. Rev. Lett.}\ }\textbf {\bibinfo
  {volume} {113}},\ \bibinfo {pages} {165503} (\bibinfo {year}
  {2014})}\BibitemShut {NoStop}%
\bibitem [{\citenamefont {Ruban}\ and\ \citenamefont
  {Razumovskiy}(2012)}]{Ruban2012}%
  \BibitemOpen
  \bibfield  {author} {\bibinfo {author} {\bibfnamefont {A.~V.}\ \bibnamefont
  {Ruban}}\ and\ \bibinfo {author} {\bibfnamefont {V.~I.}\ \bibnamefont
  {Razumovskiy}},\ }\bibfield  {title} {\bibinfo {title} {First-principles
  based thermodynamic model of phase equilibria in bcc {Fe-Cr} alloys},\ }\href
  {https://doi.org/https://doi.org/10.1103/PhysRevB.86.174111} {\bibfield
  {journal} {\bibinfo  {journal} {Phys. Rev. B}\ }\textbf {\bibinfo {volume}
  {86}},\ \bibinfo {pages} {174111} (\bibinfo {year} {2012})}\BibitemShut
  {NoStop}%
\bibitem [{\citenamefont {K{\"o}rmann}\ \emph {et~al.}(2011)\citenamefont
  {K{\"o}rmann}, \citenamefont {Dick}, \citenamefont {Hickel},\ and\
  \citenamefont {Neugebauer}}]{Kormann2011}%
  \BibitemOpen
  \bibfield  {author} {\bibinfo {author} {\bibfnamefont {F.}~\bibnamefont
  {K{\"o}rmann}}, \bibinfo {author} {\bibfnamefont {A.}~\bibnamefont {Dick}},
  \bibinfo {author} {\bibfnamefont {T.}~\bibnamefont {Hickel}},\ and\ \bibinfo
  {author} {\bibfnamefont {J.}~\bibnamefont {Neugebauer}},\ }\bibfield  {title}
  {\bibinfo {title} {Role of spin quantization in determining the thermodynamic
  properties of magnetic transition metals},\ }\href
  {https://doi.org/https://doi.org/10.1103/PhysRevB.83.165114} {\bibfield
  {journal} {\bibinfo  {journal} {Phys. Rev. B}\ }\textbf {\bibinfo {volume}
  {83}},\ \bibinfo {pages} {165114} (\bibinfo {year} {2011})}\BibitemShut
  {NoStop}%
\bibitem [{\citenamefont {Abrikosov}\ \emph {et~al.}(2016)\citenamefont
  {Abrikosov}, \citenamefont {Ponomareva}, \citenamefont {Steneteg},
  \citenamefont {Barannikova},\ and\ \citenamefont {Alling}}]{Abrikosov2016}%
  \BibitemOpen
  \bibfield  {author} {\bibinfo {author} {\bibfnamefont {I.}~\bibnamefont
  {Abrikosov}}, \bibinfo {author} {\bibfnamefont {A.}~\bibnamefont
  {Ponomareva}}, \bibinfo {author} {\bibfnamefont {P.}~\bibnamefont
  {Steneteg}}, \bibinfo {author} {\bibfnamefont {S.}~\bibnamefont
  {Barannikova}},\ and\ \bibinfo {author} {\bibfnamefont {B.}~\bibnamefont
  {Alling}},\ }\bibfield  {title} {\bibinfo {title} {Recent progress in
  simulations of the paramagnetic state of magnetic materials},\ }\href
  {https://doi.org/https://doi.org/10.1016/j.cossms.2015.07.003} {\bibfield
  {journal} {\bibinfo  {journal} {Current Opinion in Solid State and Materials
  Science}\ }\textbf {\bibinfo {volume} {20}},\ \bibinfo {pages} {85} (\bibinfo
  {year} {2016})}\BibitemShut {NoStop}%
\bibitem [{\citenamefont {K{\"o}rmann}\ \emph {et~al.}(2012)\citenamefont
  {K{\"o}rmann}, \citenamefont {Dick}, \citenamefont {Grabowski}, \citenamefont
  {Hickel},\ and\ \citenamefont {Neugebauer}}]{Kormann2012}%
  \BibitemOpen
  \bibfield  {author} {\bibinfo {author} {\bibfnamefont {F.}~\bibnamefont
  {K{\"o}rmann}}, \bibinfo {author} {\bibfnamefont {A.}~\bibnamefont {Dick}},
  \bibinfo {author} {\bibfnamefont {B.}~\bibnamefont {Grabowski}}, \bibinfo
  {author} {\bibfnamefont {T.}~\bibnamefont {Hickel}},\ and\ \bibinfo {author}
  {\bibfnamefont {J.}~\bibnamefont {Neugebauer}},\ }\bibfield  {title}
  {\bibinfo {title} {Atomic forces at finite magnetic temperatures: Phonons in
  paramagnetic iron},\ }\href
  {https://doi.org/https://doi.org/10.1103/PhysRevB.85.125104} {\bibfield
  {journal} {\bibinfo  {journal} {Phys. Rev. B}\ }\textbf {\bibinfo {volume}
  {85}},\ \bibinfo {pages} {125104} (\bibinfo {year} {2012})}\BibitemShut
  {NoStop}%
\bibitem [{\citenamefont {Hegde}\ \emph {et~al.}(2020)\citenamefont {Hegde},
  \citenamefont {Grabowski}, \citenamefont {Zhang}, \citenamefont {Waseda},
  \citenamefont {Hickel}, \citenamefont {Freysoldt},\ and\ \citenamefont
  {Neugebauer}}]{Hegde2020}%
  \BibitemOpen
  \bibfield  {author} {\bibinfo {author} {\bibfnamefont {O.}~\bibnamefont
  {Hegde}}, \bibinfo {author} {\bibfnamefont {M.}~\bibnamefont {Grabowski}},
  \bibinfo {author} {\bibfnamefont {X.}~\bibnamefont {Zhang}}, \bibinfo
  {author} {\bibfnamefont {O.}~\bibnamefont {Waseda}}, \bibinfo {author}
  {\bibfnamefont {T.}~\bibnamefont {Hickel}}, \bibinfo {author} {\bibfnamefont
  {C.}~\bibnamefont {Freysoldt}},\ and\ \bibinfo {author} {\bibfnamefont
  {J.}~\bibnamefont {Neugebauer}},\ }\bibfield  {title} {\bibinfo {title}
  {Atomic relaxation around defects in magnetically disordered materials
  computed by atomic spin constraints within an efficient {Lagrange}
  formalism},\ }\href
  {https://doi.org/https://doi.org/10.1103/PhysRevB.102.144101} {\bibfield
  {journal} {\bibinfo  {journal} {Phys. Rev. B}\ }\textbf {\bibinfo {volume}
  {102}},\ \bibinfo {pages} {144101} (\bibinfo {year} {2020})}\BibitemShut
  {NoStop}%
\bibitem [{\citenamefont {Gambino}\ and\ \citenamefont
  {Alling}(2018)}]{Alling2018}%
  \BibitemOpen
  \bibfield  {author} {\bibinfo {author} {\bibfnamefont {D.}~\bibnamefont
  {Gambino}}\ and\ \bibinfo {author} {\bibfnamefont {B.}~\bibnamefont
  {Alling}},\ }\bibfield  {title} {\bibinfo {title} {Lattice relaxations in
  disordered {Fe}-based materials in the paramagnetic state from first
  principles},\ }\href
  {https://doi.org/https://doi.org/10.1103/PhysRevB.98.064105} {\bibfield
  {journal} {\bibinfo  {journal} {Phys. Rev. B}\ }\textbf {\bibinfo {volume}
  {98}},\ \bibinfo {pages} {064104} (\bibinfo {year} {2018})}\BibitemShut
  {NoStop}%
\bibitem [{\citenamefont {Hegde}\ \emph {et~al.}(2021)\citenamefont {Hegde},
  \citenamefont {Kulitckii}, \citenamefont {Schneider}, \citenamefont
  {Soisson}, \citenamefont {Hickel}, \citenamefont {Neugebauer}, \citenamefont
  {Wilde}, \citenamefont {Divinski},\ and\ \citenamefont {Fu}}]{Hegde2021}%
  \BibitemOpen
  \bibfield  {author} {\bibinfo {author} {\bibfnamefont {O.}~\bibnamefont
  {Hegde}}, \bibinfo {author} {\bibfnamefont {V.}~\bibnamefont {Kulitckii}},
  \bibinfo {author} {\bibfnamefont {A.}~\bibnamefont {Schneider}}, \bibinfo
  {author} {\bibfnamefont {F.}~\bibnamefont {Soisson}}, \bibinfo {author}
  {\bibfnamefont {T.}~\bibnamefont {Hickel}}, \bibinfo {author} {\bibfnamefont
  {J.}~\bibnamefont {Neugebauer}}, \bibinfo {author} {\bibfnamefont
  {G.}~\bibnamefont {Wilde}}, \bibinfo {author} {\bibfnamefont
  {S.}~\bibnamefont {Divinski}},\ and\ \bibinfo {author} {\bibfnamefont
  {C.-C.}\ \bibnamefont {Fu}},\ }\bibfield  {title} {\bibinfo {title} {Impact
  of magnetic transition on {Mn} diffusion in $\alpha$–iron: Correlative
  state-of-the-art theoretical and experimental study},\ }\href
  {https://doi.org/https://doi.org/10.1103/PhysRevB.104.184107} {\bibfield
  {journal} {\bibinfo  {journal} {Phys. Rev. B}\ }\textbf {\bibinfo {volume}
  {104}},\ \bibinfo {pages} {184107} (\bibinfo {year} {2021})}\BibitemShut
  {NoStop}%
\bibitem [{\citenamefont {Schneider}\ \emph {et~al.}(2020)\citenamefont
  {Schneider}, \citenamefont {Fu}, \citenamefont {Soisson},\ and\ \citenamefont
  {Barreteau}}]{Schneider2020}%
  \BibitemOpen
  \bibfield  {author} {\bibinfo {author} {\bibfnamefont {A.}~\bibnamefont
  {Schneider}}, \bibinfo {author} {\bibfnamefont {C.-C.}\ \bibnamefont {Fu}},
  \bibinfo {author} {\bibfnamefont {F.}~\bibnamefont {Soisson}},\ and\ \bibinfo
  {author} {\bibfnamefont {C.}~\bibnamefont {Barreteau}},\ }\bibfield  {title}
  {\bibinfo {title} {Atomic diffusion in $\alpha$-iron across the {Curie}
  point: An efficient and transferable ab initio–based modeling approach},\
  }\href {https://doi.org/https://doi.org/10.1103/PhysRevLett.124.215901}
  {\bibfield  {journal} {\bibinfo  {journal} {Phys. Rev. Lett.}\ }\textbf
  {\bibinfo {volume} {124}},\ \bibinfo {pages} {215901} (\bibinfo {year}
  {2020})}\BibitemShut {NoStop}%
\bibitem [{\citenamefont {Acet}\ \emph {et~al.}(1994)\citenamefont {Acet},
  \citenamefont {Zähres}, \citenamefont {Wassermann},\ and\ \citenamefont
  {Pepperhoff}}]{Acet1994}%
  \BibitemOpen
  \bibfield  {author} {\bibinfo {author} {\bibfnamefont {M.}~\bibnamefont
  {Acet}}, \bibinfo {author} {\bibfnamefont {H.}~\bibnamefont {Zähres}},
  \bibinfo {author} {\bibfnamefont {E.~F.}\ \bibnamefont {Wassermann}},\ and\
  \bibinfo {author} {\bibfnamefont {W.}~\bibnamefont {Pepperhoff}},\ }\bibfield
   {title} {\bibinfo {title} {High-temperature moment-volume instability and
  anti-{Invar} of {$\gamma$-Fe}},\ }\href
  {https://doi.org/https://doi.org/10.1103/PhysRevB.49.6012} {\bibfield
  {journal} {\bibinfo  {journal} {Phys. Rev. B}\ }\textbf {\bibinfo {volume}
  {49}},\ \bibinfo {pages} {6012} (\bibinfo {year} {1994})}\BibitemShut
  {NoStop}%
\bibitem [{\citenamefont {Vineyard}(1957)}]{Vineyard1957}%
  \BibitemOpen
  \bibfield  {author} {\bibinfo {author} {\bibfnamefont {G.~H.}\ \bibnamefont
  {Vineyard}},\ }\bibfield  {title} {\bibinfo {title} {Frequency factors and
  isotope effects in solid state rate processes},\ }\href
  {https://doi.org/https://doi.org/10.1016/0022-3697(57)90059-8} {\bibfield
  {journal} {\bibinfo  {journal} {J. Phys. Chem. Solids}\ }\textbf {\bibinfo
  {volume} {3}},\ \bibinfo {pages} {121} (\bibinfo {year} {1957})}\BibitemShut
  {NoStop}%
\bibitem [{\citenamefont {Garnier}\ \emph {et~al.}(2013)\citenamefont
  {Garnier}, \citenamefont {Manga}, \citenamefont {Trinkle}, \citenamefont
  {Nastar},\ and\ \citenamefont {Bellon}}]{Garnier2013}%
  \BibitemOpen
  \bibfield  {author} {\bibinfo {author} {\bibfnamefont {T.}~\bibnamefont
  {Garnier}}, \bibinfo {author} {\bibfnamefont {V.~R.}\ \bibnamefont {Manga}},
  \bibinfo {author} {\bibfnamefont {D.~R.}\ \bibnamefont {Trinkle}}, \bibinfo
  {author} {\bibfnamefont {M.}~\bibnamefont {Nastar}},\ and\ \bibinfo {author}
  {\bibfnamefont {P.}~\bibnamefont {Bellon}},\ }\bibfield  {title} {\bibinfo
  {title} {Stress-induced anisotropic diffusion in alloys: Complex {Si} solute
  flow near a dislocation core in {Ni}},\ }\href
  {https://doi.org/http://dx.doi.org/10.1103/PhysRevB.88.134108} {\bibfield
  {journal} {\bibinfo  {journal} {Phys. Rev. B}\ }\textbf {\bibinfo {volume}
  {88}},\ \bibinfo {pages} {134108} (\bibinfo {year} {2013})}\BibitemShut
  {NoStop}%
\bibitem [{\citenamefont {Tapasa}\ \emph {et~al.}(2006)\citenamefont {Tapasa},
  \citenamefont {Barashev}, \citenamefont {Bacon},\ and\ \citenamefont
  {Osetsky}}]{Osetsky2006}%
  \BibitemOpen
  \bibfield  {author} {\bibinfo {author} {\bibfnamefont {K.}~\bibnamefont
  {Tapasa}}, \bibinfo {author} {\bibfnamefont {A.~V.}\ \bibnamefont
  {Barashev}}, \bibinfo {author} {\bibfnamefont {D.~J.}\ \bibnamefont
  {Bacon}},\ and\ \bibinfo {author} {\bibfnamefont {Y.~N.}\ \bibnamefont
  {Osetsky}},\ }\bibfield  {title} {\bibinfo {title} {Computer simulation of
  carbon diffusion and vacancy–carbon interaction in {$\alpha$}-iron},\
  }\href {https://doi.org/https://doi.org/10.1016/j.actamat.2006.05.029}
  {\bibfield  {journal} {\bibinfo  {journal} {Acta Materialia}\ }\textbf
  {\bibinfo {volume} {55}},\ \bibinfo {pages} {1–11} (\bibinfo {year}
  {2006})}\BibitemShut {NoStop}%
\bibitem [{\citenamefont {Ma}\ and\ \citenamefont {Dudarev}(2016)}]{Ma2016}%
  \BibitemOpen
  \bibfield  {author} {\bibinfo {author} {\bibfnamefont {P.-W.}\ \bibnamefont
  {Ma}}\ and\ \bibinfo {author} {\bibfnamefont {S.~L.}\ \bibnamefont
  {Dudarev}},\ }\bibfield  {title} {\bibinfo {title} {Spin waves and
  {Heisenberg} exchange constants for {$\alpha$}-iron},\ }\href
  {https://doi.org/http://dx.doi.org/10.1103/PhysRevB.91.054420} {\bibfield
  {journal} {\bibinfo  {journal} {Phys. Rev. B}\ }\textbf {\bibinfo {volume}
  {172}},\ \bibinfo {pages} {054420} (\bibinfo {year} {2016})}\BibitemShut
  {NoStop}%
\bibitem [{\citenamefont {Kresse}\ and\ \citenamefont
  {Hafner}(1993)}]{Kresse1993}%
  \BibitemOpen
  \bibfield  {author} {\bibinfo {author} {\bibfnamefont {G.}~\bibnamefont
  {Kresse}}\ and\ \bibinfo {author} {\bibfnamefont {J.}~\bibnamefont
  {Hafner}},\ }\bibfield  {title} {\bibinfo {title} {Ab initio molecular
  dynamics for liquid metals},\ }\href
  {https://doi.org/https://doi.org/10.1103/PhysRevB.47.558} {\bibfield
  {journal} {\bibinfo  {journal} {Phys. Rev. B}\ }\textbf {\bibinfo {volume}
  {47}},\ \bibinfo {pages} {558} (\bibinfo {year} {1993})}\BibitemShut
  {NoStop}%
\bibitem [{\citenamefont {Kresse}\ and\ \citenamefont
  {Hafner}(1994)}]{Kresse1994}%
  \BibitemOpen
  \bibfield  {author} {\bibinfo {author} {\bibfnamefont {G.}~\bibnamefont
  {Kresse}}\ and\ \bibinfo {author} {\bibfnamefont {J.}~\bibnamefont
  {Hafner}},\ }\bibfield  {title} {\bibinfo {title} {Ab initio
  molecular{-}dynamics simulation of the
  liquid{-}metal{-}amorphous{-}semiconductor transition in germanium},\ }\href
  {https://doi.org/https://doi.org/10.1103/PhysRevB.49.14251} {\bibfield
  {journal} {\bibinfo  {journal} {Phys. Rev. B}\ }\textbf {\bibinfo {volume}
  {49}},\ \bibinfo {pages} {14251} (\bibinfo {year} {1994})}\BibitemShut
  {NoStop}%
\bibitem [{\citenamefont {Kresse}\ and\ \citenamefont
  {Furthm{\"u}ller}(1996{\natexlab{a}})}]{Kresse1996}%
  \BibitemOpen
  \bibfield  {author} {\bibinfo {author} {\bibfnamefont {G.}~\bibnamefont
  {Kresse}}\ and\ \bibinfo {author} {\bibfnamefont {J.}~\bibnamefont
  {Furthm{\"u}ller}},\ }\bibfield  {title} {\bibinfo {title} {Efficiency of
  ab{-}initio total energy calculations for metals and semiconductors using a
  plane{-}wave basis set},\ }\href
  {https://doi.org/https://doi.org/10.1016/0927-0256(96)00008-0} {\bibfield
  {journal} {\bibinfo  {journal} {Comput. Mat. Sci.}\ }\textbf {\bibinfo
  {volume} {6}},\ \bibinfo {pages} {15} (\bibinfo {year}
  {1996}{\natexlab{a}})}\BibitemShut {NoStop}%
\bibitem [{\citenamefont {Kresse}\ and\ \citenamefont
  {Furthm{\"u}ller}(1996{\natexlab{b}})}]{Kresse1996B}%
  \BibitemOpen
  \bibfield  {author} {\bibinfo {author} {\bibfnamefont {G.}~\bibnamefont
  {Kresse}}\ and\ \bibinfo {author} {\bibfnamefont {J.}~\bibnamefont
  {Furthm{\"u}ller}},\ }\bibfield  {title} {\bibinfo {title} {Efficient
  iterative schemes for ab initio total{-}energy calculations using a
  plane{-}wave basis set},\ }\href
  {https://doi.org/https://doi.org/10.1103/PhysRevB.54.11169} {\bibfield
  {journal} {\bibinfo  {journal} {Phys. Rev. B}\ }\textbf {\bibinfo {volume}
  {54}},\ \bibinfo {pages} {11169} (\bibinfo {year}
  {1996}{\natexlab{b}})}\BibitemShut {NoStop}%
\bibitem [{\citenamefont {Kresse}\ and\ \citenamefont
  {Joubert}(1999)}]{Kresse1999}%
  \BibitemOpen
  \bibfield  {author} {\bibinfo {author} {\bibfnamefont {G.}~\bibnamefont
  {Kresse}}\ and\ \bibinfo {author} {\bibfnamefont {D.}~\bibnamefont
  {Joubert}},\ }\bibfield  {title} {\bibinfo {title} {From ultrasoft
  pseudopotentials to the projector augmented wave method},\ }\href
  {https://doi.org/https://doi.org/10.1103/PhysRevB.59.1758} {\bibfield
  {journal} {\bibinfo  {journal} {Phys. Rev. B}\ }\textbf {\bibinfo {volume}
  {59}},\ \bibinfo {pages} {1758} (\bibinfo {year} {1999})}\BibitemShut
  {NoStop}%
\bibitem [{\citenamefont {Potter}(1934)}]{Potter1934}%
  \BibitemOpen
  \bibfield  {author} {\bibinfo {author} {\bibfnamefont {H.~H.}\ \bibnamefont
  {Potter}},\ }\bibfield  {title} {\bibinfo {title} {The magneto-caloric effect
  and other magnetic phenomena in iron},\ }\href
  {https://doi.org/https://doi.org/10.1098/rspa.1934.0161} {\bibfield
  {journal} {\bibinfo  {journal} {Proceedings of the Royal Society of London
  A}\ }\textbf {\bibinfo {volume} {146}},\ \bibinfo {pages} {362} (\bibinfo
  {year} {1934})}\BibitemShut {NoStop}%
\bibitem [{\citenamefont {Crangle}\ and\ \citenamefont
  {Goodman}(1971)}]{Crangle1971}%
  \BibitemOpen
  \bibfield  {author} {\bibinfo {author} {\bibfnamefont {J.}~\bibnamefont
  {Crangle}}\ and\ \bibinfo {author} {\bibfnamefont {G.~M.}\ \bibnamefont
  {Goodman}},\ }\bibfield  {title} {\bibinfo {title} {The magnetization of pure
  iron and nickel},\ }\href
  {https://doi.org/https://doi.org/10.1098/rspa.1971.0044} {\bibfield
  {journal} {\bibinfo  {journal} {Proceedings of the Royal Society of London
  A}\ }\textbf {\bibinfo {volume} {321}},\ \bibinfo {pages} {477} (\bibinfo
  {year} {1971})}\BibitemShut {NoStop}%
\bibitem [{\citenamefont {Lundow}\ \emph {et~al.}(2009)\citenamefont {Lundow},
  \citenamefont {Markström},\ and\ \citenamefont {Rosengren}}]{Lundow2009}%
  \BibitemOpen
  \bibfield  {author} {\bibinfo {author} {\bibfnamefont {P.~H.}\ \bibnamefont
  {Lundow}}, \bibinfo {author} {\bibfnamefont {K.}~\bibnamefont {Markström}},\
  and\ \bibinfo {author} {\bibfnamefont {A.}~\bibnamefont {Rosengren}},\
  }\bibfield  {title} {\bibinfo {title} {The ising model for the bcc, fcc and
  diamond lattices: A comparison},\ }\href
  {https://doi.org/https://doi.org/10.1080/14786430802680512} {\bibfield
  {journal} {\bibinfo  {journal} {Phil. Mag.}\ }\textbf {\bibinfo {volume}
  {89}},\ \bibinfo {pages} {2009} (\bibinfo {year} {2009})}\BibitemShut
  {NoStop}%
\bibitem [{\citenamefont {Arajs}\ and\ \citenamefont
  {Miller}(1960)}]{Arajs1960}%
  \BibitemOpen
  \bibfield  {author} {\bibinfo {author} {\bibfnamefont {S.}~\bibnamefont
  {Arajs}}\ and\ \bibinfo {author} {\bibfnamefont {D.~S.}\ \bibnamefont
  {Miller}},\ }\bibfield  {title} {\bibinfo {title} {Paramagnetic
  susceptibilities of {Fe} and {Fe–Si} alloys},\ }\href
  {https://doi.org/https://doi.org/10.1063/1.1735788} {\bibfield  {journal}
  {\bibinfo  {journal} {J. Appl. Phys.}\ }\textbf {\bibinfo {volume} {31}},\
  \bibinfo {pages} {986} (\bibinfo {year} {1960})}\BibitemShut {NoStop}%
\bibitem [{\citenamefont {Stukowski}(2010)}]{Stukowski2010}%
  \BibitemOpen
  \bibfield  {author} {\bibinfo {author} {\bibfnamefont {A.}~\bibnamefont
  {Stukowski}},\ }\bibfield  {title} {\bibinfo {title} {Visualization and
  analysis of atomistic simulation data with {OVITO -} the open visualization
  tool},\ }\href
  {https://doi.org/https://doi.org/10.1088/0965-0393/18/1/015012} {\bibfield
  {journal} {\bibinfo  {journal} {Model. Simul. Mater. Sci. Eng.}\ }\textbf
  {\bibinfo {volume} {18}},\ \bibinfo {pages} {015012} (\bibinfo {year}
  {2010})}\BibitemShut {NoStop}%
\bibitem [{\citenamefont {Blaiszik}\ \emph {et~al.}(2016)\citenamefont
  {Blaiszik}, \citenamefont {Chard}, \citenamefont {Pruyne}, \citenamefont
  {Ananthakrishnan}, \citenamefont {Tuecke},\ and\ \citenamefont
  {Foster}}]{Blaszik2016}%
  \BibitemOpen
  \bibfield  {author} {\bibinfo {author} {\bibfnamefont {B.}~\bibnamefont
  {Blaiszik}}, \bibinfo {author} {\bibfnamefont {K.}~\bibnamefont {Chard}},
  \bibinfo {author} {\bibfnamefont {J.}~\bibnamefont {Pruyne}}, \bibinfo
  {author} {\bibfnamefont {R.}~\bibnamefont {Ananthakrishnan}}, \bibinfo
  {author} {\bibfnamefont {S.}~\bibnamefont {Tuecke}},\ and\ \bibinfo {author}
  {\bibfnamefont {I.}~\bibnamefont {Foster}},\ }\bibfield  {title} {\bibinfo
  {title} {The materials data facility: Data services to advance materials
  science research},\ }\href
  {https://doi.org/https://doi.org/10.1007/s11837-016-2001-3} {\bibfield
  {journal} {\bibinfo  {journal} {JOM}\ }\textbf {\bibinfo {volume} {68}},\
  \bibinfo {pages} {2045–2052} (\bibinfo {year} {2016})}\BibitemShut
  {NoStop}%
\bibitem [{\citenamefont {Blaiszik}\ \emph {et~al.}(2019)\citenamefont
  {Blaiszik}, \citenamefont {Ward}, \citenamefont {Schwarting}, \citenamefont
  {Gaff}, \citenamefont {Chard}, \citenamefont {Pike}, \citenamefont {Chard},\
  and\ \citenamefont {Foster}}]{Blaszik2019}%
  \BibitemOpen
  \bibfield  {author} {\bibinfo {author} {\bibfnamefont {B.}~\bibnamefont
  {Blaiszik}}, \bibinfo {author} {\bibfnamefont {L.}~\bibnamefont {Ward}},
  \bibinfo {author} {\bibfnamefont {M.}~\bibnamefont {Schwarting}}, \bibinfo
  {author} {\bibfnamefont {J.}~\bibnamefont {Gaff}}, \bibinfo {author}
  {\bibfnamefont {R.}~\bibnamefont {Chard}}, \bibinfo {author} {\bibfnamefont
  {D.}~\bibnamefont {Pike}}, \bibinfo {author} {\bibfnamefont {K.}~\bibnamefont
  {Chard}},\ and\ \bibinfo {author} {\bibfnamefont {I.}~\bibnamefont
  {Foster}},\ }\bibfield  {title} {\bibinfo {title} {A data ecosystem to
  support machine learning in materials science},\ }\href
  {https://doi.org/https://doi.org/10.1557/mrc.2019.118} {\bibfield  {journal}
  {\bibinfo  {journal} {MRS Commun.}\ }\textbf {\bibinfo {volume} {9}},\
  \bibinfo {pages} {1125–1133} (\bibinfo {year} {2019})}\BibitemShut
  {NoStop}%
\bibitem [{\citenamefont {Wirth}\ and\ \citenamefont
  {Trinkle}(2025)}]{Wirth2025data}%
  \BibitemOpen
  \bibfield  {author} {\bibinfo {author} {\bibfnamefont {L.~J.}\ \bibnamefont
  {Wirth}}\ and\ \bibinfo {author} {\bibfnamefont {D.~R.}\ \bibnamefont
  {Trinkle}},\ }\href {https://doi.org/https://doi.org/10.18126/fttq-w045}
  {\bibinfo {title} {Data citation: {S}pin-space averaged {Fe–C} diffusion
  calculations from density functional theory}} (\bibinfo {year}
  {2025})\BibitemShut {NoStop}%
\bibitem [{\citenamefont {Perdew}\ \emph {et~al.}(1996)\citenamefont {Perdew},
  \citenamefont {Burke},\ and\ \citenamefont {Ernzerhof}}]{Perdew1996}%
  \BibitemOpen
  \bibfield  {author} {\bibinfo {author} {\bibfnamefont {J.~P.}\ \bibnamefont
  {Perdew}}, \bibinfo {author} {\bibfnamefont {K.}~\bibnamefont {Burke}},\ and\
  \bibinfo {author} {\bibfnamefont {M.}~\bibnamefont {Ernzerhof}},\ }\bibfield
  {title} {\bibinfo {title} {Generalized gradient approximation made simple},\
  }\href {https://doi.org/https://doi.org/10.1103/PhysRevLett.77.3865}
  {\bibfield  {journal} {\bibinfo  {journal} {Phys. Rev. Lett.}\ }\textbf
  {\bibinfo {volume} {77}},\ \bibinfo {pages} {3865} (\bibinfo {year}
  {1996})}\BibitemShut {NoStop}%
\bibitem [{\citenamefont {Blöchl}(1994)}]{Blochl1994}%
  \BibitemOpen
  \bibfield  {author} {\bibinfo {author} {\bibfnamefont {P.~E.}\ \bibnamefont
  {Blöchl}},\ }\bibfield  {title} {\bibinfo {title} {Projector augmented-wave
  method},\ }\href {https://doi.org/https://doi.org/10.1103/PhysRevB.50.17953}
  {\bibfield  {journal} {\bibinfo  {journal} {Phys. Rev. B}\ }\textbf {\bibinfo
  {volume} {50}},\ \bibinfo {pages} {17953} (\bibinfo {year}
  {1994})}\BibitemShut {NoStop}%
\bibitem [{\citenamefont {Methfessel}\ and\ \citenamefont
  {Paxton}(1989)}]{Methfessel1989}%
  \BibitemOpen
  \bibfield  {author} {\bibinfo {author} {\bibfnamefont {M.}~\bibnamefont
  {Methfessel}}\ and\ \bibinfo {author} {\bibfnamefont {A.~T.}\ \bibnamefont
  {Paxton}},\ }\bibfield  {title} {\bibinfo {title} {High{-}precision sampling
  for {B}rillouin{-}zone integration in metals},\ }\href
  {https://doi.org/https://doi.org/10.1103/PhysRevB.40.3616} {\bibfield
  {journal} {\bibinfo  {journal} {Phys. Rev. B}\ }\textbf {\bibinfo {volume}
  {40}},\ \bibinfo {pages} {3616} (\bibinfo {year} {1989})}\BibitemShut
  {NoStop}%
\bibitem [{\citenamefont {Fellinger}\ \emph {et~al.}(2017)\citenamefont
  {Fellinger}, \citenamefont {Louis G.~Hector},\ and\ \citenamefont
  {Trinkle}}]{Fellinger2017}%
  \BibitemOpen
  \bibfield  {author} {\bibinfo {author} {\bibfnamefont {M.~R.}\ \bibnamefont
  {Fellinger}}, \bibinfo {author} {\bibfnamefont {J.}~\bibnamefont {Louis
  G.~Hector}},\ and\ \bibinfo {author} {\bibfnamefont {D.~R.}\ \bibnamefont
  {Trinkle}},\ }\bibfield  {title} {\bibinfo {title} {Ab initio calculations of
  the lattice parameter and elastic stiffness coefficients of bcc {Fe} with
  solutes},\ }\href
  {https://doi.org/http://dx.doi.org/10.1016/j.commatsci.2016.09.040}
  {\bibfield  {journal} {\bibinfo  {journal} {Comp. Mater. Sci.}\ }\textbf
  {\bibinfo {volume} {126}},\ \bibinfo {pages} {503} (\bibinfo {year}
  {2017})}\BibitemShut {NoStop}%
\bibitem [{\citenamefont {Monkhorst}\ and\ \citenamefont
  {Pack}(1976)}]{Monkhorst1976}%
  \BibitemOpen
  \bibfield  {author} {\bibinfo {author} {\bibfnamefont {H.~J.}\ \bibnamefont
  {Monkhorst}}\ and\ \bibinfo {author} {\bibfnamefont {J.~D.}\ \bibnamefont
  {Pack}},\ }\bibfield  {title} {\bibinfo {title} {Special points for
  {B}rillouin{-}zone integrations},\ }\href
  {https://doi.org/https://doi.org/10.1103/PhysRevB.13.5188} {\bibfield
  {journal} {\bibinfo  {journal} {Phys. Rev. B}\ }\textbf {\bibinfo {volume}
  {13}},\ \bibinfo {pages} {5188} (\bibinfo {year} {1976})}\BibitemShut
  {NoStop}%
\end{thebibliography}

%
\newpage


\section*{Supplementary material}

# External magnetic field suppression of carbon diffusion in iron: Supplemental material

Luke J. Wirth and Dallas R. Trinkle[*]

*Department of Materials Science and Engineering,*

*University of Illinois, Urbana-Champaign, Illinois 61801, USA*

(Dated: July 21, 2025)

## Abstract

Supplemental material: (1) magnetic model methodology and parameterization; (2) DFT parameters and magnetic moment constraint details; (3) sampling statistics and temperature dependencies of spin-space averaged (SSA) activation energies; (4) Zeeman splitting energy calculation details and results.

## S1. MAGNETIC MODEL DETAILS

To generate spin spaces, we use Monte Carlo simulations of a Heisenberg model in a large bcc Fe supercell to reproduce experimentally observed net magnetizations at high temperatures with and without external magnetic fields. Local 3D magnetic moment configurations from these serve as input to noncollinear DFT calculations as constraint directions for moments of Fe atoms. Our model uses the Heisenberg Hamiltonian

$$H = -J \sum_{k,l} \mathbf{M}_k \cdot \mathbf{M}_l - \sum_k \mathbf{B}_{\text{ext}} \cdot \mathbf{M}_k , \quad \text{(S1)}$$

where $J$ is an exchange interaction parameter, $k$ denotes the index of each Fe atom with magnetic moment $\mathbf{M}$, and $l$ values are the indices of the nearest neighbors of atom $k$. Cells are periodic with side lengths $L = 32$ and contain a total of 65,536 atoms. Trial moves see the direction of a random individual moment randomly oriented and accepted if $-k_B T \times \ln(r) > \Delta H$, where $r$ is a random number drawn from the interval $[0, 1)$ and $\Delta H$ is the change in Eq. S1 induced by the trial move. Beginning with moments of unit magnitude, we vary $T$ and $J$ to identify $J = 43.2$ meV as yielding the empirical $T_C$ of 1043 K for bcc Fe. Here, $T_C$ is defined as the temperature where susceptibility $\frac{\partial M}{\partial T}$ in our cell is at a maximum. Next, we fix $(J|\mathbf{M}|)^2$ and adjust $|\mathbf{M}|$ until the net

---

[*] dtrinkle@illinois.edu

1

magnetizations in our cell agree with experimental observations under 0.5, 1.0, and 1.5 T fields over temperature ranges near $T_C$ [32]. Fig. S1 shows that setting $|\mathbf{M}| = 2.79$ $\mu_B$ leads to good agreement at all three fields. After tuning, we run the simulations and extract local 54-moment samples that are separated from one another by at least the autocorrelation time calculated based on the net magnetization in the cell. At 1043 K, 0 T, where we expect this time to be the longest, separations of 1,000 Monte Carlo sweeps are more than enough to ensure sample independence.

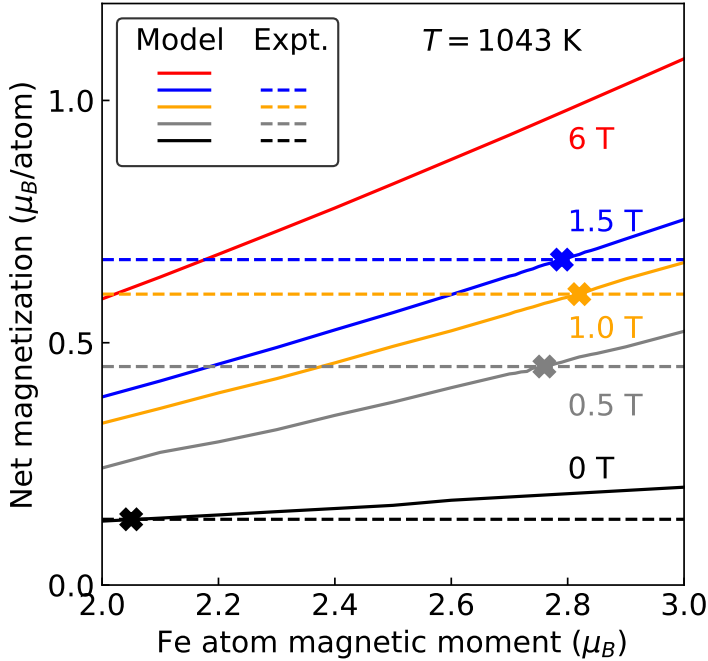

FIG. S1. Comparison of how net magnetization in Monte Carlo simulations at 1043 K varies with iron atom magnetic moment and field strength. Horizontal dashed lines indicate experimentally observed magnetizations at zero-field [31] and with field [32], where × symbols indicate model magnetic moment magnitudes that yield the targeted experimental values.

## S2. DFT CALCULATION DETAILS

DFT calculations use the Vienna Ab initio Simulation Package (VASP) 5.4.4 [26–29] with the Perdew-Burke-Ernzerhof (PBE) [39] generalized gradient approximation (GGA) exchange-correlation functional and projector-augmented wave (PAW) potentials [30, 40]. Pseudopotentials for carbon and iron respectively have electronic configurations [He]$2s^2 2p^2$ and [Ar]$3d^7 4s^1$ with

maximum plane wave energies of 400.0 and 267.882 meV [30]. Calculations use plane wave cutoff energies of 550 meV, with order-one Methfessel-Paxton smearing [41] and a smearing width of 0.2 eV, following our earlier work on solutes in bcc Fe [42]. Noncollinear (NCL) calculations here use $6 \times 6 \times 6$ Monkhorst-Pack [43] $k$-meshes for Fe$_{54}$ and Fe$_{54}$C supercells.

Magnetic moment direction constraints use the algorithm of Ma and Dudarev [25], which is officially supported in VASP 6.4.0+ but implemented here by modification of the VASP source file `constrmag.F`. NCL calculations use a self-consistent loop tolerance of $10^{-4}$ eV, a penalty weight $\lambda$ of 30 eV for enforcement of constraints, and PAW radii (0.76 Å for carbon and 1.395 Å for iron) for integration of magnetic moments within spheres. A general recommendation for using VASP with constrained moments is to converge sequential calculations with increasingly large $\lambda$ penalties to encourage stability. Here, that typically resulted in new convergence difficulties upon each increase, leading to the decision to converge one batch at 30 eV for each step of the atomic relaxation. Most magnetic configuration calculations could eventually converge at a given set of $(T, B)$ conditions for all three of the Fe$_{54}$, octahedral Fe$_{54}$C, and tetrahedral Fe$_{54}$C geometries. A few particularly troublesome sets of moment directions needed to be replaced: one set at (1043 K, 6 T), two at (986 K, 0 T), and one in the disordered local moment (DLM) case.

## S3. TEMPERATURE DEPENDENCE AND STATISTICS OF ENERGY BARRIERS

Fig. S2 illustrates the distribution of energy barriers $Q$ at each set of conditions used for spin-space averaging, with and without correctional terms applied to account for the effects of thermal expansion. Most magnetic configurations see greater compression when carbon is at the octahedral rather than tetrahedral site, so volumetric thermal expansion tends to stabilize the octahedral site more, resulting in an increase of barrier with temperature. The FM barrier is 0.86 eV without correction and 0.91 eV at 1043 K, which is a similar increase as that seen by the SSA barriers at 986 and 1043 K. The DLM case sees more pronounced shifts in activation energies with temperature because it sees a greater difference in stress experienced in the octahedral than tetrahedral case. As discussed in the main text, these corrections adjust the 0 K volume supercell energies by $\frac{1}{3}\text{Tr}(\overline{P}_{ij})\frac{V(T)}{V(0\,\text{K})}$. Experimental lattice constants come from a work by Acet, et al. [21]. Because this reference reports a continuous set of observations from near 0 K to the ferrite-austenite transition temperature, it is particularly well suited as a source for estimating $\frac{V(T)}{V(0\,\text{K})}$ ratios. For example, the volume of bcc Fe is ~4.0% greater at 1043 K than at 0 K, and ~3.7% greater at 986 K. We

compute standard errors as SE = $\frac{s}{\sqrt{n}}$, where $s$ is the sample standard deviation and $n$ is the sample size. The 25 energies calculated at each set of conditions come from geometric relaxations that accounted for symmetry, so error calculations use $n = 25 \times 8$ for octahedral energies (where there are eight symmetry operations available) and $n = 25 \times 4$ for energies of the tetrahedral transition state (where there are four). Standard errors of differences are from standard deviations of the 25 sampled differences computed at each magnetic configuration, with $n = 25 \times 4$ from the tetrahedral case used to produce an upper bound on our standard error value.

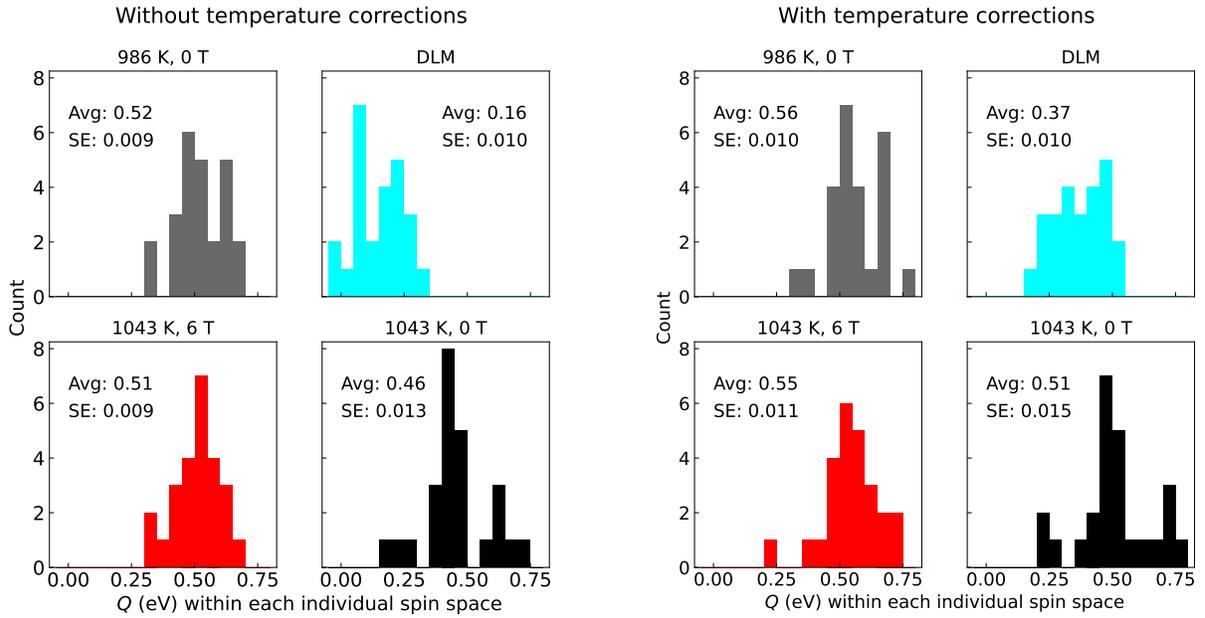

FIG. S2. Distributions of energy barriers $Q$ to carbon hops in bcc Fe within individual magnetic environments at each studied set of conditions, along with averages and standard errors (SE). Energies in the left set of panels are the unadjusted differences in $Fe_{54}C$ supercell energies when carbon is at the transition state relative to the octahedral site. The right set of panels includes corrections that account for the effects of thermal expansion on the barrier. DLM corrections use the 1043 K volume; as discussed in the main text, this demonstrates that the average DLM barrier is too low to describe diffusion experiments at 1043 K.

## S4. ZEEMAN SPLITTING ENERGY CALCULATIONS AT 0 K

Fig. S3 illustrates the field magnitudes that would be necessary to significantly affect the activation energy barrier to carbon diffusion in ferromagnetic bcc Fe at 0 K, demonstrating that splitting energy effects do not explain the observed suppression of diffusion by fields $\lesssim 10$ T. Magnitudes $B$

4

[T] relate to splitting energies $\Delta E_B$ [eV] according to $B = -\frac{g}{2}\mu_B \Delta E_B$, where $g$ is the Landé g-factor and $\mu_B$ is the Bohr magneton. Each activation energy here is the difference between energies when carbon is at a tetrahedral transition state relative to the octahedral equilibrium site. Collinear (CL) spin-polarized calculations used VASP `bext.F` routines to apply the splitting energy as a difference in energy between up and down spin states. These used most of the same settings as our NCL calculations, but because of the reduced computational demands of CL calculations, took place in $Fe_{128}C$ supercells using a $10^{-8}$ eV self-consistent loop threshold, still with $6 \times 6 \times 6$ k-meshes. The ferromagnetic barrier converged to 0.862 eV with NCL calculation parameters and 0.866 eV with CL parameters, indicating good agreement between the two sets of settings.

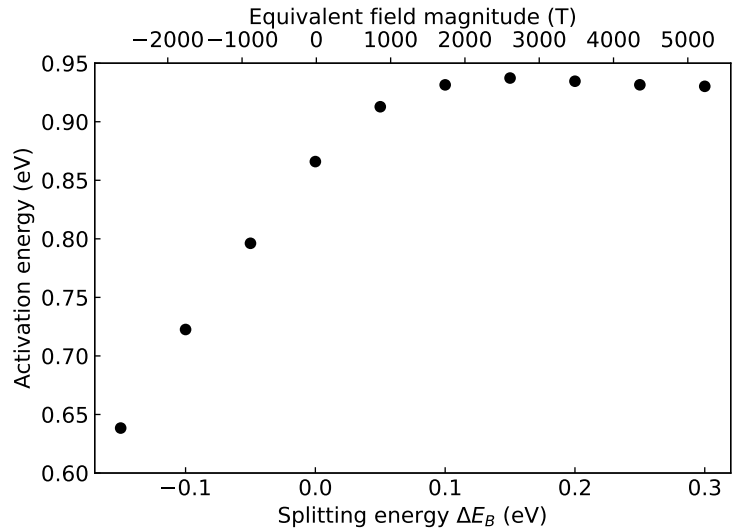

FIG. S3. Response of the activation energy barrier of carbon diffusion in bcc Fe to Zeeman splitting energies that represent external magnetic fields at 0 K. Positive field magnitudes align with collinear bcc Fe moments while negative magnitudes oppose them. At splitting energies below –0.15 eV, the moments spontaneously flip to align with the field.

[1] J. R. G. da Silva and R. B. McLellan, Diffusion of carbon and nitrogen in b.c.c. iron, Mater. Sci. and Eng. **26**, 83 (1976).

[2] J. K. Stanley, The diffusion and solubility of carbon in alpha iron, JOM **1**, 752 (1949).

[3] H. Fujii and S. Tsurekawa, Diffusion of carbon in iron under magnetic fields, Phys. Rev. B **83**, 054412 (2011).

[4] M. S. Kesler, M. J. Thompson, D. Weiss, and M. V. Manuel, Simultaneously improving process efficiency and mechanical properties in aluminum alloys with applied magnetic fields, in *Light Metals 2025*, edited by L. Edwards (Springer Nature Switzerland, Cham, 2025) p. 246–251.

[5] L. Ruch, D. R. Sain, H. L. Yeh, and L. Girifalco, Analysis of diffusion in ferromagnets, J. Phys. Chem. Solids **37**, 649 (1976).

[6] R. B. McLellan, M. L. Rudee, and T. Ishibachi, The thermodynamics of dilute interstitial solid solutions with dual-site occupancy and its application to the diffusion of carbon in alpha iron, AIME Met. Soc. Trans. **233**, 1938 (1965).

[7] D. E. Jiang and E. A. Carter, Carbon dissolution and diffusion in ferrite and austenite from first principles, Phys. Rev. B **67**, 214103 (2003).

[8] C. Domain, C. S. Becquart, and J. Foct, Ab initio study of foreign interstitial atom (C,N) interactions with intrinsic point defects in $\alpha$–Fe, Phys. Rev. B **69**, 144112 (2004).

[9] L. A. Girifalco, Activation energy for diffusion in ferromagnetics, J. Phys. Chem. Solids **23**, 1171 (1962).

[10] M. Wuttig, On the activation entropy of interstitial diffusion in b.c.c. iron, Scripta Metall. **5**, 33 (1971).

[11] R. Farraro and R. B. McLellan, The diffusion of heavy interstitial solute atoms in body-centered cubic metals, Mater. Sci. and Eng. **39**, 47–56 (1978).

[12] C. A. Wert, Diffusion coefficient of C in $\alpha$-iron, Phys. Rev. **79**, 601 (1950).

[13] E. Bousquet, N. A. Spaldin, and K. T. Delaney, Unexpectedly large electronic contribution to linear magnetoelectricity, Phys. Rev. Lett. **106**, 107202 (2011).

[14] F. Körmann, B. Grabowski, B. Dutta, T. Hickel, B. Fultz, and J. Neugebauer, Temperature dependent magnon-phonon coupling in bcc Fe from theory and experiment, Phys. Rev. Lett. **113**, 165503 (2014).

[15] F. Körmann, A. Dick, T. Hickel, and J. Neugebauer, Role of spin quantization in determining the thermodynamic properties of magnetic transition metals, Phys. Rev. B **83**, 165114 (2011).

[16] F. Körmann, A. Dick, B. Grabowski, T. Hickel, and J. Neugebauer, Atomic forces at finite magnetic temperatures: Phonons in paramagnetic iron, Phys. Rev. B **85**, 125104 (2012).

[17] O. Hegde, M. Grabowski, X. Zhang, O. Waseda, T. Hickel, C. Freysoldt, and J. Neugebauer, Atomic relaxation around defects in magnetically disordered materials computed by atomic spin constraints within an efficient Lagrange formalism, Phys. Rev. B **102**, 144101 (2020).

[18] D. Gambino and B. Alling, Lattice relaxations in disordered Fe-based materials in the paramagnetic state from first principles, Phys. Rev. B **98**, 064104 (2018).

[19] O. Hegde, V. Kulitckii, A. Schneider, F. Soisson, T. Hickel, J. Neugebauer, G. Wilde, S. Divinski, and C.-C. Fu, Impact of magnetic transition on Mn diffusion in $\alpha$–iron: Correlative state-of-the-art theoretical and experimental study, Phys. Rev. B **104**, 184107 (2021).

[20] A. Schneider, C.-C. Fu, F. Soisson, and C. Barreteau, Atomic diffusion in $\alpha$-iron across the Curie point: An efficient and transferable ab initio–based modeling approach, Phys. Rev. Lett. **124**, 215901 (2020).

[21] M. Acet, H. Zähres, E. F. Wassermann, and W. Pepperhoff, High-temperature moment-volume instability and anti-Invar of $\gamma$-Fe, Phys. Rev. B **49**, 6012 (1994).

[22] G. H. Vineyard, Frequency factors and isotope effects in solid state rate processes, J. Phys. Chem. Solids **3**, 121 (1957).

[23] T. Garnier, V. R. Manga, D. R. Trinkle, M. Nastar, and P. Bellon, Stress-induced anisotropic diffusion in alloys: Complex Si solute flow near a dislocation core in Ni, Phys. Rev. B **88**, 134108 (2013).

[24] K. Tapasa, A. V. Barashev, D. J. Bacon, and Y. N. Osetsky, Computer simulation of carbon diffusion and vacancy–carbon interaction in $\alpha$-iron, Acta Materialia **55**, 1–11 (2006).

[25] P.-W. Ma and S. L. Dudarev, Spin waves and Heisenberg exchange constants for $\alpha$-iron, Phys. Rev. B **172**, 054420 (2016).

[26] G. Kresse and J. Hafner, Ab initio molecular dynamics for liquid metals, Phys. Rev. B **47**, 558 (1993).

[27] G. Kresse and J. Hafner, Ab initio molecular-dynamics simulation of the liquid-metal-amorphous-semiconductor transition in germanium, Phys. Rev. B **49**, 14251 (1994).

[28] G. Kresse and J. Furthmüller, Efficiency of ab-initio total energy calculations for metals and semiconductors using a plane-wave basis set, Comput. Mat. Sci. **6**, 15 (1996).

[29] G. Kresse and J. Furthmüller, Efficient iterative schemes for ab initio total-energy calculations using a plane-wave basis set, Phys. Rev. B **54**, 11169 (1996).

[30] G. Kresse and D. Joubert, From ultrasoft pseudopotentials to the projector augmented wave method, Phys. Rev. B **59**, 1758 (1999).

[31] H. H. Potter, The magneto-caloric effect and other magnetic phenomena in iron, Proceedings of the Royal Society of London A **146**, 362 (1934).

[32] J. Crangle and G. M. Goodman, The magnetization of pure iron and nickel, Proceedings of the Royal Society of London A **321**, 477 (1971).

[33] P. H. Lundow, K. Markström, and A. Rosengren, The ising model for the bcc, fcc and diamond lattices: A comparison, Phil. Mag. **89**, 2009 (2009).

[34] S. Arajs and D. S. Miller, Paramagnetic susceptibilities of Fe and Fe–Si alloys, J. Appl. Phys. **31**, 986 (1960).

[35] A. Stukowski, Visualization and analysis of atomistic simulation data with OVITO - the open visualization tool, Model. Simul. Mater. Sci. Eng. **18**, 015012 (2010).

[36] B. Blaiszik, K. Chard, J. Pruyne, R. Ananthakrishnan, S. Tuecke, and I. Foster, The materials data facility: Data services to advance materials science research, JOM **68**, 2045–2052 (2016).

[37] B. Blaiszik, L. Ward, M. Schwarting, J. Gaff, R. Chard, D. Pike, K. Chard, and I. Foster, A data ecosystem to support machine learning in materials science, MRS Commun. **9**, 1125–1133 (2019).

[38] L. J. Wirth and D. R. Trinkle, Data citation: Spin-space averaged Fe–C diffusion calculations from density functional theory (2025).

[39] J. P. Perdew, K. Burke, and M. Ernzerhof, Generalized gradient approximation made simple, Phys. Rev. Lett. **77**, 3865 (1996).

[40] P. E. Blöchl, Projector augmented-wave method, Phys. Rev. B **50**, 17953 (1994).

[41] M. Methfessel and A. T. Paxton, High-precision sampling for Brillouin-zone integration in metals, Phys. Rev. B **40**, 3616 (1989).

[42] M. R. Fellinger, J. Louis G. Hector, and D. R. Trinkle, Ab initio calculations of the lattice parameter and elastic stiffness coefficients of bcc Fe with solutes, Comp. Mater. Sci. **126**, 503 (2017).

[43] H. J. Monkhorst and J. D. Pack, Special points for Brillouin-zone integrations, Phys. Rev. B **13**, 5188 (1976).


\end{document}